\documentclass[
letterpaper,
showpacs,floatfix, aps,prb,amsmath,amssymb,
twocolumn,
groupaddress,eqsecnum]{revtex4}

\usepackage{graphicx}

\newcommand{\be}{\begin{equation}}
\newcommand{\ee}{\end{equation}}
\newcommand{\bea}{\begin{eqnarray}}
\newcommand{\eea}{\end{eqnarray}}
\newcommand*{\bs}{\begin{split}}
\newcommand*{\bes}{\begin{equation}\begin{split}}
\newcommand*{\ees}{\end{split} \end{equation}}
\newcommand*{\nl}{\nonumber\\}
\newcommand*{\tn}{|t_n|^2}
\newcommand*{\nc}{N_{ch}}
\newcommand*{\sumc}{\sum_{n=1}^{\nc}}

\newcommand*{\ghat}{\hat{g}_n}
\newcommand*{\Ghat}{\hat{g}}
\newcommand*{\fhat}{\hat{f}_n}

\newcommand*{\one}{\hat{1}}

\newcommand*{\fk}{f_n^K}
\newcommand*{\gk}{g^K}
\newcommand*{\ff}{\mathcal{F}}
\newcommand*{\D}{\mathcal{D}}
\newcommand*{\K}{\mathcal{K}}
\newcommand*{\C}{\mathcal{C}}

\renewcommand*{\Im}{\mathrm{Im}}
\renewcommand*{\Re}{\mathrm{Re}} 
\newcommand*{\Tr}{\mathrm{Tr}}
\newcommand*{\glr}{\frac{1}{\tau}}

\newcommand*{\coh}[1]{\coth{\left(\frac{#1}{4T}\right)}}
\newcommand*{\ipsi}[3]{\chi_{#1}^{#2,#3}(\omega)}
\newcommand*{\J}{\mathcal{J}(\epsilon,\omega)}
\newcommand*{\myxi}[3]{\xi_{#1}^{#2,#3}}
\newcommand*{\myxic}[3]{\xi_{#1}^{#2,-#3}}
\newcommand*{\rmg}{\mathrm{g}}
\newcommand*{\rmh}{\mathrm{h}}
\newcommand*{\St}{\widehat{\mathrm{St}}}
\newcommand*{\Stb}{\widehat{\boldsymbol{\mathrm{St}}}}
\newcommand*{\Li}{\mathrm{Li}_2}
\newcommand{\req}[1]{Eq.~(\ref{#1})}
\newcommand{\reqs}[1]{Eqs.~(\ref{#1})}

\begin{document}

\title{Spin Related Effects in Transport Properties of ``Open'' Quantum Dots
}

\author{Y. Ahmadian, G. Catelani and I.L. Aleiner}
\affiliation{Physics Department, Columbia University, New York, NY 10027}


\begin{abstract}

We study the interaction corrections to the transport coefficients in open 
quantum dots (i.e. dots connected to leads of large conductance $G \gg e^2/\pi\hbar$), via a quantum kinetic equation approach. The effects of all the 
channels of the universal (in the Random Matrix Theory sense) interaction 
Hamiltonian are accounted for at one loop approximation. For the electrical conductance we find that even though the magnitude of the 
triplet channel interaction is smaller than the charging energy, the 
differential conductance at small bias is greatly affected by this interaction.
Furthermore, the application of a magnetic field can significantly change the
conductance due to the Zeeman splitting, producing finite bias anomalies. For the thermal conductance we find 
that the Wiedemann-Franz law is violated by the interaction corrections, and we investigated the effect of magnetic field on the Lorentz ratio for contacts of finite reflection. The charge and triplet 
channel corrections to the electrical and thermal conductance vanish for reflectionless contacts.  In the latter case the temperature and magnetic field dependence of the conductance is determined by the Maki-Thompson correction in the Cooper channel.

\end{abstract}
\date{\today}
\maketitle

\section{Introduction}
\label{intro}

Transport properties of quantum dots are strongly affected by electron-electron
interactions, the most studied example being the so-called Coulomb blockade
phenomenon\cite{zeller:giaver,kulik:shekhter,averin:likharev} (see Ref.~\onlinecite{aleiner:brouwer:glazman:02} for a review) in 
dots connected to external electrodes 
by contacts of low conductance, $G \ll e^2/\pi\hbar \equiv G_q$: if the
temperature $T$ is smaller than the charging energy $E_c$ associated with 
the addition of one electron to the dot, transport through the dot is 
exponentially suppressed (except at charge degeneracy points).
As the contacts' conductance increases, quantum fluctuations of the dot's
charge will eventually lift the Coulomb blockade; even so, interactions in
such ``open'' ($G \gg G_q$)  or weakly blockaded dots do influence transport 
phenomena.\cite{brouwer:aleiner:99,golubev:zaikin:04,brouwer:lamackraft:05} 
In this paper we consider the interaction corrections to transport coefficients
of first order in $1/\rmg \equiv G_q/G$, similar to 
the well-known corrections in the bulk.
\cite{altshuler:aronov:book}

\begin{figure}[!t]
\includegraphics[height=115pt,width=170pt,angle=0]{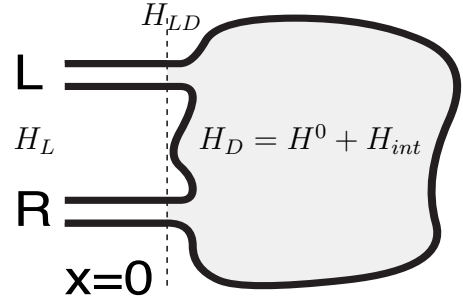}
\large
\put(-100,63){$H_D=H^0 + H_{int}$}
\put(-122,110){$H_{LD}$}
\put(-170,63){$H_L$}
\caption{Quantum dot connected to two leads, and the corresponding Hamiltonians.}
\label{figdot}
\end{figure}

It is known that transport properties of dots with a large number of electrons can be described within the
framework of Random Matrix Theory (RMT) (see e.g. Refs.~\onlinecite{beenakker:97,aleiner:brouwer:glazman:02,alhassid} for reviews). We consider a system separated into clean leads without interaction, a closed dot and the coupling between the dot and the leads as shown in Fig.~\ref{figdot}. All interference and interaction effects are associated with the closed dot Hamiltonian $H_D$. In the RMT 
approach, the orbital dynamics of non-interacting spinless electrons in the closed dot is described by the 
$M \rightarrow \infty$ limit of an $M\times M$ Hamiltonian matrix 
with random entries $H^0_{mn}$ belonging to the Gaussian ensemble:
\be\label{RMT}
\langle H^0_{mn} H^0_{ij} \rangle_{_{RM}} = M\frac{\delta_1^2}{\pi^2}
\Big[\delta_{mj}\delta_{ni} + \left(1-\frac{\rmg_h}{4M}\right)\delta_{mi}\delta_{nj}\Big]\, ,
\ee
where $\delta_1$ is the one-particle mean level spacing, $\langle\cdots\rangle_{_{RM}}$ denotes averaging over the random matrix ensemble, and $\rmg_h$ is a dimensionless parameters quantifying the orbital effect of the weak magnetic field. This can be estimated as:
\be\label{ghdef}
\rmg_h\simeq \frac{E_T}{\delta_1}(\Phi/\Phi_0)^2, 
\ee
where $E_T$ is the Thouless energy\cite{aleiner:brouwer:glazman:02} 
(above which scale the RMT description is not applicable), $\Phi$ is the total magnetic flux through the dot, and $\Phi_0=hc/e$ is
the flux quantum. The extreme values, $\rmg=0$ and $\rmg=4M$, correspond to the orthogonal and unitary ensembles respectively, while intermediate values describe the crossover between those.

\noindent

In the absence of spin-orbit interactions (we neglect spin-orbit interactions in this paper; the interplay of exchange and spin-orbit interactions was studied in Ref.~\onlinecite{gorokhov:brouwer:03}), the spin of the electrons can be accounted for by defining the following non-interacting Hamiltonian operator (in the second quantized notation):
\be
H^0 = \psi^\dagger_m H^0_{mn} \psi_n,
\ee
where $\psi_{m}^\dagger$ $(\psi_m)$ is a two component spinor operator, whose components create (annihilate) electrons in the $m$-th orbital state, with specific spin projections. For 
interacting electrons, the closed dot Hamiltonian is given by 
\be
H_D = H^0 + H_{int},
\ee
where the dominant part of the interaction Hamiltonian has the universal form:\cite{kurland:al:al:04,aleiner:brouwer:glazman:02}
\be\label{hint}
H_{int} =  E_c N^2 + J_s \vec{S}^2 + J_c \mathcal{T}^{\dagger}\mathcal{T}.
\ee
Here
\begin{subequations}
\bea
 N &=& \sum_{n=1}^{M} \psi^{\dagger}_n \psi_n,\label{nhat}\\
\vec{S} &=& \frac{1}{2}\sum_{n=1}^{M} \psi^{\dagger}_n \vec{\sigma} \psi_n,\label{spin}\\
\mathcal{T} &=& \frac{1}{2}\sum_{n=1}^{M} \psi_n \sigma^y \psi_n,\label{T}
\eea\end{subequations}
are respectively the number of electrons in the dot, the total spin of the dot electrons, and the pairing operator. Here, $ \vec{\sigma}$ and $\sigma^y$ are Pauli matrices acting in the spinor space.

The last term in the right hand side of \req{hint} describes 
pairing between electrons (interaction in the Cooper channel)
and for $J_c<0$ it drives the dot towards the superconducting state; studies of such 
superconducting grains have been reviewed e.g. in  Ref.~\onlinecite{vondelft:01}.
Here we will 
assume that either $J_c> 0$ and hence no superconducting transition at any temperature, or $J_c<0$ but $T-T_c\gtrsim T_c$, so that the dot is in the normal phase, and furthermore, the superconducting fluctuations are small.
The second term (triplet channel) gives the dependence of the dot's energy
on the total spin in the dot -- the effects of this term on the tunneling
density of states and on the spin susceptibility\cite{gefen:05} 
and on the peak spacing\cite{vorojtsov:05} in the Coulomb blockade 
regime have been recently considered.
Finally the first term (singlet channel)
describes the charging energy and it is responsible for the Coulomb blockade. 
In the weakly blockaded regime, only this term has been considered previously 
in the 
literature,\cite{brouwer:aleiner:99,golubev:zaikin:04,brouwer:lamackraft:05}
as its contribution is expected to be the dominant one (for the repulsive 
Coulomb interaction). However this term is 
not affected by an external 
magnetic field, whereas the two remaining terms are. Our goal is
to calculate the interaction corrections to the transport coefficients
with the full universal Hamiltonian $H_D$ taken into account
and to examine  the dependence of these corrections on the applied magnetic 
field. In particular, we consider the non-linear conductance for 
voltage-biased dots and their (linear response) thermal conductance. To 
evaluate these transport coefficients we construct, starting from the RMT
description, a quantum kinetic equation analogous to the one developed for
the description of disordered metals.\cite{catelani:aleiner:05,zala:n:a:01}

The remainder of the paper is organized as follows: in the next section we summarize
our results for the conductance of metallic quantum dots. In Sec.
\ref{sec1} we  present the derivation of the 0-dimensional
Usadel equation in RMT. This equation is the starting point for the derivation 
of the kinetic equation as outlined in Sec.~\ref{sec:kineq}. In 
Sec.~\ref{sec:calc} we give the explicit calculation of the transport 
coefficients.

\section{Summary of the results}
\label{sec:results}

Here we present our results for the interaction corrections to the 
differential electrical conductance and the linear thermal conductance of 
quantum dots in the presence of a magnetic field. These results are derived in 
Sec.~\ref{sec:calc}. 

\subsection{Electrical Conductance}

The total ensemble averaged differential conductance $G$ is
\be
G =\frac{d I}{d V}=  G_0 + \Delta G,
\ee
where 
\be\label{clcon}
G_0=\frac{e^2}{\pi\hbar}\frac{\rmg_L \rmg_R}{\rmg_L + \rmg_R}
\ee
is the classical conductance and $\Delta G$ is the interaction correction. We do not include the weak localization correction for non-interacting electrons which can be found e.g. in Ref.~\onlinecite{beenakker:97}. We also do not study the contribution of $\Delta G$ to the mesoscopic fluctuations of the conductance, which are smaller than $\Delta G$ by the additional factor $G_q/G_0$, (For the charge channel this effect was studied in Ref.~\onlinecite{brouwer:lamackraft:05}). 

In \req{clcon} $\rmg_{L}$ and $\rmg_R$ are the dimensionless 
conductances of the left and right contacts respectively:
\be\label{glgr}
\rmg_L= \sum_{n=1}^{N_L} T_n \, , \quad
\rmg_R = \sum_{n=N_L+1}^{N_{ch}} T_n,
\ee
where $T_n$ is the transmission coefficient of the $n$-th channel.

The interaction correction $\Delta G$ has distinct contributions from 
each term in the interaction Hamiltonian (\ref{hint}):
\be
\Delta G = \Delta G_c + \Delta G_s +\Delta G_{Cooper}.
\ee 
We postpone the discussion of the Cooper channel correction, $\Delta G_{Cooper}$, until the end of this section. For the charge and triplet channel contributions we have found
\be\label{condu}
\Delta G_{c}=\frac{e^2}{\pi\hbar} \frac{\rmh_L \rmg_R^2 + \rmh_R \rmg_L^2}{(\rmg_L + \rmg_R)^3} \,\Xi\left(\Gamma_0,4E_c,T;V\right),
\ee
and 
\bes\label{tripl}
\Delta G_{s} = &\frac{e^2}{\pi\hbar} \frac{\rmh_L \rmg_R^2 + \rmh_R \rmg_L^2}{(\rmg_L + \rmg_R)^3} \\
&\times \sum_{m=0,\pm 1}  \Xi\left(\Gamma_0 + i m E_Z^*, J_s,T;V\right).
\end{split}
\ee
The form factors $\rmh_L$and $\rmh_R$ are given by
\be\label{hlr}
\rmh_{L(R)} =  \sum_{n \in L(R)} T_n(1-T_n). 
 \ee
Factors of this form were first obtained in Ref.~\onlinecite{matveev:y:g:93} and reproduced in Refs.~\onlinecite{brouwer:lamackraft:05,golubev:zaikin:04}. This structure was originally missed in the formalism of Ref.~\onlinecite{brouwer:aleiner:99} but was recovered in Ref.~\onlinecite{brouwer:lamackraft:05}.

In \reqs{condu} and (\ref{tripl}) the dimensionless function $\Xi$ is defined by
\be \begin{split}
&\Xi\left(\Gamma,\mathcal{E},T;V\right) \equiv \\ & \Re \frac{\Gamma_0}{ \Gamma} \left[ \Psi\left(\frac{\Gamma -  i e V}{2\pi T}\right)-\Psi\left(\frac{(1+\frac{ \mathcal{E}}{\delta_1}) \Gamma - i e V}{2\pi T}\right) \right],
\end{split}
\ee
where
\be
\Psi(z) \equiv \psi^{(0)}(z) + z \psi^{(1)}(z),
\ee
and $\psi^{(i)}(z)$ is the $i$-th derivative of the digamma function. 
The form of this function agrees with the result derived in Refs.~\onlinecite{golubev:zaikin:04,brouwer:lamackraft:05} for the charge channel. We denote 
with $V$ and $T$ the bias voltage and the temperature respectively, 
$\Gamma_0/\hbar$ is the escape rate:
\be\label{gma0}
\Gamma_0 \equiv \frac{\hbar}{\tau}= \frac{ \delta_1}{2\pi}\sum_{n=1}^{\nc} T_n,
\ee 
and $E_Z^*$ is the Zeeman energy renormalized by the exchange interaction:
\be\label{EZ}
E_Z^* = \frac{E_Z}{1+J_s/\delta_1} = \frac{g_L\mu_B B}{1+J_s/\delta_1}.
\ee
Here $B$ is the magnetic field, $g_L$ is the Lande g-factor, and $\mu_B$ is the Bohr magneton.

\begin{figure}[!bt]
\includegraphics[height=2.4in,width=3.5in,angle=0]{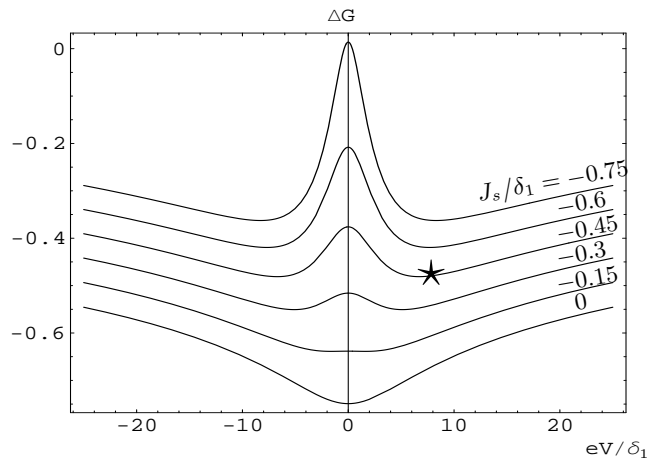}
\put(-70,98){\rotatebox{10}{$J_s/\delta_1=-0.75$}}
\put(-40,93){\rotatebox{10}{$-0.6$}}
\put(-40,83){\rotatebox{10}{$-0.45$}}
\put(-40,74){\rotatebox{10}{$-0.3$}}
\put(-40,64){\rotatebox{10}{$-0.15$}}
\put(-33,57){\rotatebox{10}{$0$}}
\put(-92,66){\huge $\star$}
\put(-8,0.8){\tiny $1$}
\caption{Interaction correction to the dimensionless differential conductance  for varying strength of the triplet interaction in the absence of Zeeman splitting. Here $T = 0.1 \delta_1$, $\Gamma_0 = 5 \delta_1$ and $E_c = 100 \delta_1$  where $\delta_1$ is the mean level spacing, and we have taken $N_L=N_R$ and $T_n=1/2$ for all channels. The value of $J_s$ is shown above each graph. For clarity, graphs are shifted upwards by $0.05$ at each step.}
\label{fig1}
\end{figure}

\begin{figure}[tbh]
\includegraphics[height=2.4in,width=3.5in,angle=0]{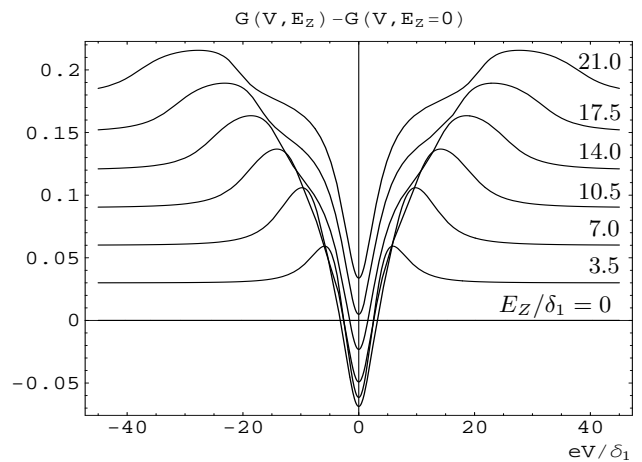}
\put(-59,57){$E_Z/\delta_1 = 0$}
\put(-25,72){$ 3.5$}
\put(-25,86){$ 7.0$}
\put(-29,100){$ 10.5$}
\put(-29,115){$ 14.0$}
\put(-29,130){$ 17.5$}
\put(-29,149){$ 21.0$}
\put(-12.5,1){\tiny $1$}
\caption{The magneto-conductance of the starred curve of Fig.~\ref{fig1} vs. bias voltage , for different values of Zeeman splitting energy (shown above each graph).  The graphs are shifted upwards by $0.03$ at each step. }
\label{fig2}
\end{figure}

\begin{figure}[tbh]
\includegraphics[height=2.4in,width=3.5in,angle=0]{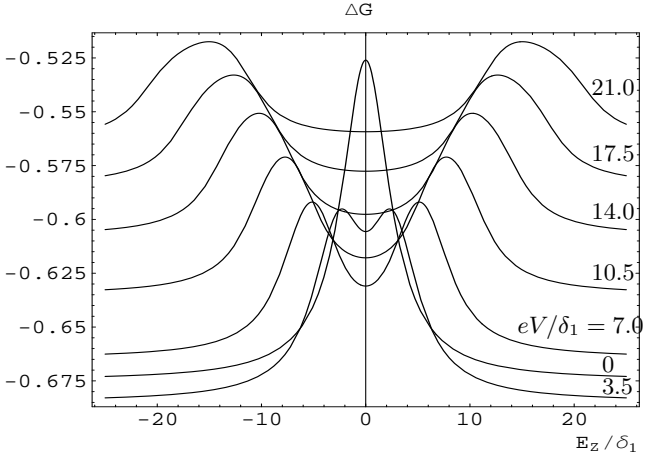}
\put(-52,46){$eV/\delta_1 = 7.0$}
\put(-20,23){$ 3.5$}
\put(-20,31){$ 0$}
\put(-24,66){$ 10.5$}
\put(-24,89){$ 14.0$}
\put(-24,111){$ 17.5$}
\put(-24,136){$ 21.0$}
\put(-10,1){\tiny $1$}
\caption{The magneto-conductance of the starred curve of Fig.~\ref{fig1} vs. Zeeman splitting energy, at various bias voltages (shown above each graph). The graphs are not shifted here.}
\label{fig22}
\end{figure}\noindent

In Fig.~\ref{fig1} we plot $\Delta G = \Delta G_c + \Delta G_s$ in units of $G_q = e^2/\pi\hbar$, for different values of the triplet channel interaction constant $J_s$ in the absence of magnetic field. We have taken $N_L=N_R$ and $T_n=1/2$ for all channels, so that the prefactors in \reqs{condu} and (\ref{tripl}) are equal to $1/8$.
The interaction constant takes values ranging from $J_s=0$ to 
$J_s=-0.75\delta_1$, which apply to most metals and quantum 
dots (the values of $J_s/\delta_1$ are reviewed e.g. in  Ref.~\onlinecite{gorokhov:brouwer:03}). The charging
energy $E_c$ on the other hand is much larger than the mean level spacing. A 
theoretical estimate for two dimensional dots\cite{aleiner:brouwer:glazman:02}
yields 
$E_c/\delta_1 \simeq r_s k_F L$, where $r_s$ is the gas parameter, $k_F$ is the 
Fermi wavelength, and $L$ is the lateral dimension of the dot. This is large 
because of the large factor $k_F L\gg 1$.
As shown in Fig.~\ref{fig1}, the charge channel correction generates a dip in 
the conductance at zero bias with a characteristic width of order 
$(1+4E_c/\delta_1)\Gamma_0$, analogous to the zero bias anomaly in bulk systems. 
This width can also be estimated as $\hbar/\tau_c$, where $\tau_c<\tau$ is the classical recharging time governing the charge dynamics. On the other hand, 
the attractive triplet interaction produces a peak with a 
much smaller width of order $(1+J_s/\delta_1)\Gamma_0\equiv \hbar/\tau_s$, corresponding to the slow spin dynamics: $\tau_s>\tau$. Therefore we have a 
competition between the two corrections at zero bias as seen in Fig.~\ref{fig1}.

In Fig.~\ref{fig2} we plot the magneto-conductance, $G(E_Z)-G(E_Z=0)$, for different values of the 
Zeeman energy (and with the same assumptions about the form
factors made for Fig.~\ref{fig1}). In the presence of 
a magnetic field, the triplet channel contribution decomposes into three terms 
due to the Zeeman splitting.
These terms produce peaks at $e V\simeq \pm E_Z^*$ and $0$.
As the value of the Zeeman energy increases so does the width of the displaced 
peaks (for $m=\pm1$); in the limit $E_Z\gg \Gamma_0$ this width is given by the
bare Zeeman energy $E_Z$. In Fig.~\ref{fig22} we plot $\Delta G$ vs. $E_Z$ for different bias voltages.

In the case of reflectionless contacts, ($T_n=1$ for all $n$), the corrections in \reqs{condu}
and (\ref{tripl}) vanish due to 
the vanishing of the form factors $\rmh_L$ and $\rmh_R$ of \req{hlr}; a 
non-zero contribution to $G$ is given by the zero-dimensional 
analog of the Maki-Thompson\cite{maki,thompson,larkin:80} correction to the conductivity:
\bea\label{mtcorr}
\Delta G_{Cooper}&=&\Delta G_{MT} =\left(\frac{1}{\varepsilon^2}\right) \frac{e^2}{\pi\hbar}\frac{\rmg_L \rmg_R}{(\rmg_L+\rmg_R)^4}
 \\
&&\times\sum_{\alpha,\beta = L,R} \rmg_\alpha \rmg_\beta \Upsilon_{\alpha \beta}(\Gamma_*,E_Z^*,T;V),
\nonumber \eea
where $\Upsilon_{LL}=\Upsilon_{RR}$ and $\Upsilon_{LR}=\Upsilon_{RL}$ are dimensionless functions given below, and $\Gamma_*/\hbar$ is the escape rate modified by the effect of the  the magnetic field 
on the orbital motion of the electrons [see the discussion after \req{RMT}]:
\be\label{gma1}
\Gamma_* \equiv \frac{\hbar}{\tau_*}= \frac{ \delta_1}{2\pi}\left[\sum_{n=1}^{\nc} T_n +\rmg_h\right].
\ee 
Here
\be\label{redtemp}
\varepsilon \simeq \ln\frac{T_c}{\mathrm{max}\{T,eV,\Gamma_*\}},
\ee
and $T_c$ is defined in \req{crittemp}. In the attractive case, $T_c$ is the critical 
temperature of the superconducting transition. In this case we only consider the normal state at $E_T\gg T\gg T_c$, and hence $|\varepsilon|\gg1$. In the repulsive case, $T_c\gg E_T$, and since by assumption
$E_{T}$ is larger than all the other energy scales in the problem, 
$\varepsilon \gg 1$. This means that the correction in \req{mtcorr} is logarithmically
suppressed for normal dots and is in general much smaller than the correction due to charge and triplet channels. For this reason we report here the Cooper channel correction only for reflectionless contacts, where it is the only non-vanishing correction. The results for the general 
transmission coefficient can be found in Sec.~\ref{coopercond}. 

The functions $\Upsilon_{\alpha\beta}$  cannot be calculated in a compact form and we therefore consider approximate expressions valid in certain regions of 
parameters. In the low temperature, low voltage regime
(i.e. $T , eV \ll \Gamma_*$) we have:
\be\label{poo}
\Upsilon_{\alpha\beta} = \frac{4}{\left[1 + ( E_Z^*/\Gamma_*)^2\right]^2} \left[ 
\mathcal{A}_{\alpha\beta}\left(\frac{eV}{\Gamma_*}\right)^2 + \frac{\pi^2}{3} 
\left(\frac{T}{\Gamma_*}\right)^2 \right] ,
\ee
where $\mathcal{A}_{LR}=\mathcal{A}_{RL}=1$, and $\mathcal{A}_{LL}=\mathcal{A}_{RR}=1/4$. The $T^2$ and $V^2$ dependences are due to the inelastic nature of the 
processes at the origin of the Maki-Thompson correction.\cite{varlamov:larkin}

\begin{figure}[!tb]
\includegraphics[height=2.4in,width=3.5in,angle=0]{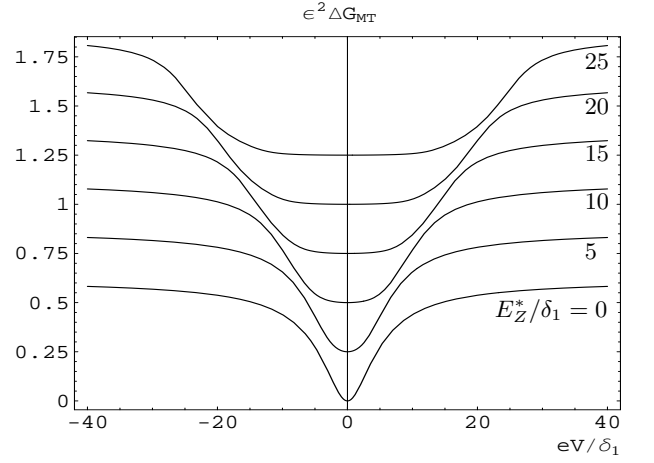}
\put(-59,53){$E_Z^*/\delta_1 = 0$}
\put(-25,75){$ 5$}
\put(-25,94){$ 10$}
\put(-25,112){$ 15$}
\put(-25,130){$ 20$}
\put(-25,147){$ 25$}
\put(-15,1){\tiny $1$}
\caption{Correction to dimensionless differential conductance for reflectionless contacts and symmetric leads at zero temperature and $\Gamma_*= 5\delta_1$. Here we have plotted $\varepsilon^2\Delta G_{MT}$, so that the actual magnitude of 
the correction is further suppressed by $1/\varepsilon^2$.
The values of the effective Zeeman splitting $E_Z^*$ are shown below each graph, and the graphs are shifted upwards by $0.25$ at each step. }
\label{fig3}
\end{figure}

At finite bias and low temperature ($T\ll eV, \Gamma_*$), we have:
\bea\label{zerot}
\Upsilon_{LR}&\approx& \frac{1}{2}\sum_{m=\pm 1} h\left(\frac{eV}{\Gamma_*},\frac{eV- m E_Z^*}{\Gamma_*}\right)\\
\Upsilon_{LL}&\approx& \frac{1}{2}\Upsilon_{LR} - \frac{1}{4}\left(\sum_{m=\pm 1} \arctan{\frac{eV-mE_Z^*}{\Gamma_*}}\right)^2,\nonumber
\eea
where the full expression for function $h$ is given in \req{hdef}. This function is used in Fig.~\ref{fig3} to plot $\varepsilon^2\Delta G_{MT}$ 
(in units of $e^2/\pi\hbar$) for symmetric leads
at different values of the Zeeman energy.
An approximate expression for $h$ is:
\bea\label{approxh} 
h & \simeq & \sum_{m,n=\pm 1}  \arctan \left( \frac{2eV + 2(m-n) E_Z^* }{\Gamma_*} 
\right) \\
&& \times \left[ \arctan \left( \frac{eV + mE_Z^*}{\Gamma_*}\right) 
- \arctan\left(\frac{mE_Z^*}{\Gamma_*}\right)\right]. \nonumber
\eea
This formula qualitatively renders the shape of the 
correction, although it overestimates it in the central plateau
[cf. Fig.\ref{fig3}];
on the other hand, \req{approxh} is a good approximation when
$|V| \gtrsim E_Z^*$ and $E_Z^* \gtrsim 1/\tau_*$, i.e. it gives a
quantitative description of both the steps at $V\sim \pm E_Z^*$ and the 
asymptotic regions at large bias for sufficiently large Zeeman energy.
At small bias ($V\ll 1/\tau_*$), the correction is better represented by
\reqs{poo}.

Finally in the high temperature regime ($T \gg \Gamma_*$) we have:
\bea
\Upsilon_{LR} &\approx& \frac{\pi^2}{2} - \frac{\pi^2}{4}\coth{\left(\frac{E_Z^*}{2T}\right)}\sum_{m=\pm }c_{1}\left(\frac{E_Z^* - m e V}{2T}\right)\nl
\Upsilon_{LL}  &\approx&  \frac{\pi^2}{8}  \sum_{m=\pm} c_{2}\left(\frac{eV-mE_Z^*}{2T}\right),
\eea
where
\be
c_{n}(x)\equiv \frac{d^n}{dx^n}(x\coth{x}).
\ee

\subsection{Thermal Conductance}\label{sec:thermcondres}

For the thermal conductance we calculate the linear response in singlet and triplet channels only, and
we do not report the
contribution of Cooper channel, as it is smaller by the factor
$1/\varepsilon^2$ [cf. \req{redtemp}].

\begin{figure}[tb]
\includegraphics[height=2.4in,width=3.5in,angle=0]{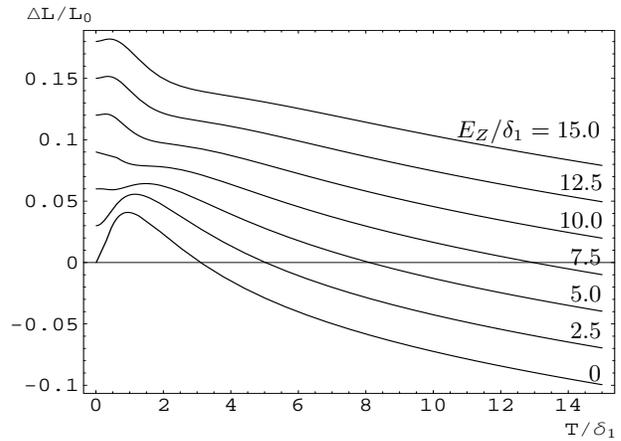}
\put(-33,26){$0$}
\put(-40,42){$ 2.5$}
\put(-40,56){$ 5.0$}
\put(-40,70){$ 7.5$}
\put(-44,84){$ 10.0$}
\put(-44,98){$ 12.5$}
\put(-83,118){$E_Z/\delta_1 =  15.0$}
\put(-26,4.5){\tiny $1$}
\caption{The relative change in the Lorentz number vs. temperature for different values of Zeeman splitting energy (shown above each graph). Here,  $ E_c = 100\delta_1$ and $J_s= -0.45\delta_1$, $\Gamma_0 = 5 \delta_1$, and we have taken $N_L=N_R=1$ and $T_n=1/2$ for all channels. The graphs are shifted upwards by $0.03$ at each step. }
\label{figther}
\end{figure}

The thermal conductance is a combination of two parts
\be\label{thcores}
\kappa = -\frac{I_L^{\varepsilon}}{T_L-T_R} =\kappa_{WF} + \Delta\kappa.
\ee 
The first term $\kappa_{WF}$ respects the Wiedemann-Franz law:
\be
\kappa_{WF}=\frac{\pi^2 T}{3e^2}  G(V=0) =\frac{\pi^2 T}{3e^2} 
 \left[G_0 +  \Delta G(V=0)\right],
\ee
with $\Delta G$ given by \reqs{condu}--(\ref{tripl}), whereas the correction
$\Delta \kappa$ violates this law:
\begin{subequations}\label{gheyn}
\be
\Delta\kappa = \Delta\kappa_c + \Delta\kappa_s,
\ee
\bes\label{kcharge}
\Delta\kappa_c = &
\frac{\rmg_R^2 \rmh_L + \rmg_L^2 \rmh_R}{(\rmg_L + \rmg_R)^3} \frac{\pi T}{9\hbar} \\
&\times  \left[
 g_1\left(\frac{2\pi T}{(1+\frac{4E_c}{\delta_1})\Gamma_0}\right) - g_1\left(\frac{2\pi T}{\Gamma_0}\right) \right],
\end{split}\ee
\bea\label{thermtripl}
&&\Delta\kappa_s =  \frac{\rmg_R^2 \rmh_L + \rmg_L^2 \rmh_R}{(\rmg_L + \rmg_R)^3} \frac{\pi T}{9\hbar}
\\ && \quad
\times\,\Re\!\!\sum_{m=0,\pm1} \frac{\Gamma_0}{\Gamma_m} \left[ g_1\left(\frac{2\pi T}{(1+\frac{J_s}{\delta_1})\Gamma_m}\right) -g_1\left(\frac{2\pi T}{\Gamma_m}\right) 
\right].
\nonumber \eea
\end{subequations}
In \reqs{gheyn} we use the notation
\be
\Gamma_m = \Gamma_0 + i m E_Z^*,
\ee
and
\be
g_1(x) = \frac{6}{x^3}\psi^{(1)}\left(\frac{1}{x}\right)-\frac{6}{x^2}-\frac{3}{x},
\ee
where $\psi^{(1)}$ is the derivative of the digamma function. The deviation from the Wiedemann-Franz law can be quantified by defining a generalized Lorentz number according to
\be
L \equiv \frac{\kappa}{T G(V=0)},
\ee 
such that it would yield the usual $L_0=\pi^2/3e^2$ in the absence of interactions. 
In Fig.~\ref{figther} we plot the relative change in the Lorentz number, $L/L_0-1$, as a function of temperature for different values of Zeeman splitting energy (and with the same assumptions about the form
factors made for Fig.~\ref{fig1}). 	
 
We note that the interaction corrections to thermal conductance \reqs{kcharge}--(\ref{thermtripl})
vanish for reflectionless contacts, and so the Wiedemann-Franz law is satisfied in this case. That this law is not violated by inelastic processes in this case can be explained by the following qualitative picture. Consider an electron with energy $\epsilon$ and charge $e$ entering the dot from the right lead with an influx proportional to the number of channels $N_R$ in this lead ($T_n =1$ for all channels). The electron will subsequently leave the dot through the right or the left lead with probability $N_R/\nc$ and $N_L/\nc$ respectively, thus contributing an outflux in the right lead proportional to $N_R^2/\nc$. Therefore, the total ingoing electric current in this lead will be
\be\label{shooz}
I_R \propto e N_R - e \frac{N_R^2}{\nc} = e \frac{N_L N_R}{N_L + N_R},
\ee
i.e. the electric conductance is given by the classical conductance \req{clcon}.
If in the process the electron loses an energy $\omega$ to the collective excitations this will still hold within the linear spectrum approximation as the bosonic excitations are neutral and do not contribute to electric current. 
This is apparently not the case for the energy current as the electron going out of the right lead now has energy $\epsilon - \omega$. However, the collective excitation with energy $\omega$ will eventually decay into an electron hole pair and these will enter the leads with the same probabilites as before, and due to energy conservation will carry the lost energy back to the right lead with the same rate, i.e. 
\be\label{birz}
I^\varepsilon_R \propto \epsilon N_R - (\epsilon - \omega) \frac{N_R^2}{\nc} - \omega \frac{N_R^2}{\nc} = \epsilon \frac{N_L N_R}{N_L + N_R}.
\ee 
We see that the relation between the energy current and the electric one \req{shooz} is not altered and we recover the Wiedemann-Franz law. 
It is crucial in this reasoning that the contacts be reflectionless, since only in this case the transmission coefficients will not be renormalized by interaction and the $T_n$'s  remain equal to one independent of energy. This is not the case for $T_n\neq 1$ and the renormalized transmission coefficients depend on energy, thus violating \req{birz} and therefore the Wiedemann-Franz law.

\section{Derivation of the 0D Usadel equation} \label{sec1}

In this section we derive the zero dimensional Usadel equation\cite{usadel}  that describes 
electrons in open quantum dots, taking into account the electron-electron
interaction but neglecting weak localization and mesoscopic fluctuation effects. 
Our treatment here is based on the Keldysh technique\cite{keldysh} 
for non-equilibrium systems,  and is analogous to 
Ref.~\onlinecite{zala:n:a:01} with appropriate modifications for the 0-d case. First we start with the description of coupling to the leads in subsection \ref{sec:leads}. In the rest of the subsections we derive equations for Green functions averaged over the appropriate random matrix ensemble. To overcome the difficulties involving ensemble averaging in the presence of the quartic interactions of the universal Hamiltonian \req{hint}, we employ a  Hubbard-Stratonovitch transformation for each channel of interaction,  rendering the fermion Hamiltonian quadratic. This is done in subsections \ref{usadel:charge} to \ref{sec:usacoop}.

\subsection{Description of the Leads}
\label{sec:leads}
The open dot [see Fig.~\ref{figdot}] 
is described by the Hamil\-tonian\cite{aleiner:brouwer:glazman:02}
\be
H = H_D + H_L + H_{LD},
\ee
where $H_D=H^0+H_{int}$ is the (RMT) Hamiltonian 
for the interacting electrons 
in the dot [see \reqs{RMT}-(\ref{hint})], 
$H_L$ is the Hamiltonian for the free electrons in the leads and 
$H_{LD}$ describes the coupling between the dot and the leads. 
The electron spectrum in the leads near the Fermi surface can be linearized: 
\be\label{hlead}
H_L = v_F \sum_{\alpha} 
\int \frac{dk}{2\pi}
 k \psi^\dagger_\alpha (k) \psi_\alpha(k),
\ee
where $v_F$ is the Fermi velocity, and $\alpha$ labels different channels in the leads: $1\leq \alpha \leq N_L$ for the
left, and $N_L+1 \leq \alpha \leq \nc$  for the 
right lead channels, and $\nc = N_L + N_R$. Here, the field $\psi_\alpha$ is understood to be a two component spinor and we suppress the spin indices unless stated otherwise. 

The leads-dot coupling Hamiltonian is 
\be\label{coupling}
H_{LD} = \sum_{\alpha, n,k} e^{-\eta \frac{|k|}{2}}(W_{n\alpha} 
\psi^\dagger_\alpha (k) \psi_n + h.c.),
\ee
where the coupling constants are defined as
\be\label{dubya}
 W_{\alpha n}  = t_n \sqrt{\frac{M \delta_1}{\pi^2 \nu}} \delta_{\alpha n},
\ee 
for $ n=\alpha \leq \nc$ and zero otherwise, 
and the exponential at $\eta \!\to\! 0^+$, is used to regularize the
coupling at large $|k|$. Here $\nu=1/(2\pi v_F)$ is  the density of states  per spin at the Fermi level.  We can always write the matrix $W$ in the above
diagonal form by choosing the appropriate basis for the random
matrix.

We introduce the exact Green functions of the electrons
in the dot (${\cal \hat G}$) and in the leads ($\hat{F}$). 
As usual they are $2\times 2$
matrices in the Keldysh and spin spaces 
 \be
\label{green}
 \begin{split}
 &\hat{\mathcal{G}}_{nm} = \left( \begin{array}{cc }
       \mathcal{G}^R_{nm}(t_1;t_2) &  \mathcal{G}^K_{nm}(t_1;t_2)  \\
     \mathcal{G}^Z_{nm}(t_1;t_2)   & \mathcal{G}^A_{nm}(t_1;t_2) 
 \end{array} \right)_K,
 \\
 &\hat{F}_{\alpha\beta}= 
\left( \begin{array}{cc }
       F^R_{\alpha\beta} (t_1,k_1;t_2,k_2) &
      F^K_{\alpha\beta} (t_1,k_1;t_2,k_2)  \\
    F_{\alpha\beta}^Z (t_1,k_1;t_2,k_2) & 
 F^A_{\alpha\beta} (t_1,k_1;t_2,k_2) 
 \end{array} \right)_K,
\end{split}
 \ee
where $1\leq n,m \leq M$
($M\to \infty$ being the size of the random matrix),
and 
$\alpha,\beta$ label different channels in each lead.
The entries of the matrix in \req{green} are given by
\bea\label{rak}
F^R & = 
& -i \theta_{12}\langle \psi_\alpha(k_1,t_1)\psi^\dagger_\beta(k_2,t_2) + 
\psi^\dagger_\beta(k_2,t_2) \psi_\alpha(k_1,t_1)\rangle\! ,\nl
F^A & = & i 
\theta_{21}\langle \psi_\alpha(k_1,t_1)\psi^\dagger_\beta(k_2,t_2) + 
\psi^\dagger_\beta(k_2,t_2) \psi_\alpha(k_1,t_1)\rangle, \nl
 F^K  & = & 
-i \langle \psi_\alpha(k_1,t_1)\psi^\dagger_\beta(k_2,t_2) -
\psi^\dagger_\beta(k_2,t_2) \psi_\alpha(k_1,
t_1)\rangle, \nl
F^Z  & = &0.
\eea
The expression for $\mathcal{G}$ is obtained from \req{rak} by replacing $\alpha,\beta$ with
$n,m$ and removing the $k$ variables. Here $\theta_{12}\equiv \theta(t_1-t_2)$, where $\theta(t)$ is the step
function, the fermionic spinor operators are in the Heisenberg representaton
and the averaging is performed over a non-equilibrium state of the
system. 

For $H_{LD}=0$, one easily finds the Green functions for the electrons
in the leads $\hat F =\hat{F}^{(0)}$
\bea
&&\hat
F^{(0)}=
2\pi\delta_{\alpha\beta}\delta(k_1-k_2)1_s\int\frac{d\epsilon}{2\pi}e^{-i
  \epsilon (t_1-t_2)} \hat{F}_\alpha^{(0)}(\epsilon,k);
\nonumber \\
&& \left[{F}_\alpha^{(0)}(\epsilon,k)\right]^R=
\frac{1}{\epsilon-v_F k+i0};
\nonumber \\
&& \left[{F}_\alpha^{(0)}(\epsilon,k)\right]^A=
\frac{1}{\epsilon-v_F k-i0};
\nonumber \\
&& \left[{F}_\alpha^{(0)}(\epsilon,k)\right]^K=
-\pi i\delta\left(\epsilon-v_F k\right) f^K_\alpha(\epsilon),
\label{f0}
\eea
where $1_s$ is the unit matrix in the spin space, and  $-2<f^K(\epsilon)<2$ 
is an arbitrary function related to the occupation 
number, $n(\epsilon)$, of the state with energy $\epsilon$ by 
$n(\epsilon)=\frac{1}{2}-\frac{1}{4}f(\epsilon)$.
For the equilibrium  Fermi distribution, $f(\epsilon)$ is given by
\be
f^K_\alpha(\epsilon) =  2 \tanh\left(\frac{\epsilon-\mu_\alpha}{2 T_\alpha}
\right). 
 \label{fkeq}
\ee
In the presence of coupling with the dot,we can use the standard
diagrammatic technique [see Fig.~\ref{fig5}.d] to write the following
expression for  $\hat F$
\bea
&&\hat{F}_{\alpha\beta}(k_1,k_2) =
\hat{F}^{(0)}_{\alpha,\beta}(k_1,k_2) + \int\frac{dk_1 ' dk_2 '}{(2\pi)^2}e^{-\eta(|k_1 '|+|k_2 '|)}
\nonumber \\&& \quad \times \, 
\hat{F}^{(0)}_{\alpha\alpha'}(k_1,k_1 ')W_{\alpha '
  n}\hat{\mathcal{G}}_{nm}W_{m\beta '}^\dagger \hat{F}^{(0)}_{\beta '
  \beta}(k_2 ',k_2),
\label{ff0}
\eea
where summation over repeated indices is implied. We generally regard the Green functions as 
operators in the time as well as in the Keldysh and spin spaces, 
so that their products are understood as operator (matrix)
multiplication in these spaces, and therefore we omit time arguments in particular.

Equation (\ref{ff0}) enables us to analyze the transport properties of the system, e.g. the electric current. 
The linearized spectrum of the leads Hamiltonian \req{hlead} suggests 
a similar linearization for the current operator in the channel $\alpha$
(we choose the direction such that outgoing current is positive)
\be
\hat{I}_{\alpha}(x)=\frac{i e}{2m}\psi^\dagger_\alpha(x)
\partial_x\psi_\alpha(x) +h.c..
\label{definition}
\ee
Substituting $\psi_\alpha(x)=\int_{-k_F}^{\infty}\frac{dk}{\pi} \psi_{\alpha}(k)
\sin(k+k_F)x$, $x<0$ into \req{definition}, and keeping the terms
not oscillating on the scale of the Fermi wavelength $2\pi/k_F$, we find (for $x=\eta_1$)
\bes
\hat{I}_{\alpha} =  \frac{- i e v_F}{2}\! 
\sum_{\alpha\in a} \int
\frac{dk dq \sin \eta_{1} q }{\pi^2}  
 \psi^\dagger_\alpha (k) \psi_\alpha(k+q)
,\end{split}
\label{current1}
\ee
where $\eta_1\to +0, \eta_1 \gg \eta$, and integration is performed for $|k|,|q| \ll k_F$. Therefore, according to \req{rak} we can write the expectation value of the outgoing current in each lead in terms of the Keldysh Green function
\be
I_a(t) = -e v_F \sum_{\alpha\in a} \int
\frac{dk dq \sin \eta_{1} q }{(2\pi)^2} \Tr_s F^K(t,k+q;t,q), 
\ee
where $a=L,R$ for the left or right lead, $\alpha$ is summed over different channels in the corresponding lead, and $\Tr_s$ is trace in the spin space. After substituting \req{f0} into \req{ff0}, to obtain $F^K$, and integrating over momenta we obtain
\be\label{tofu}
I_a  =   - \frac{ i e M \delta_1}{2\pi}  \sum_{n \in a}\tn \Tr_s\left(2 \mathcal{G}^K_{nn} - \mathcal{G}^R_{nn}.\fk + \fk . \mathcal{G}^A_{nn}\right).
\ee
Here the first term in the parentheses has the meaning of the current from the dot to the leads, and the other terms express the current from the leads to the dot. Note however, that the terms taken separately do not have any physical significance. 


\subsection{Charge Channel}
\label{usadel:charge}

\begin{figure}[!tb]
\includegraphics[height=190pt,width=236.25pt,angle=0]{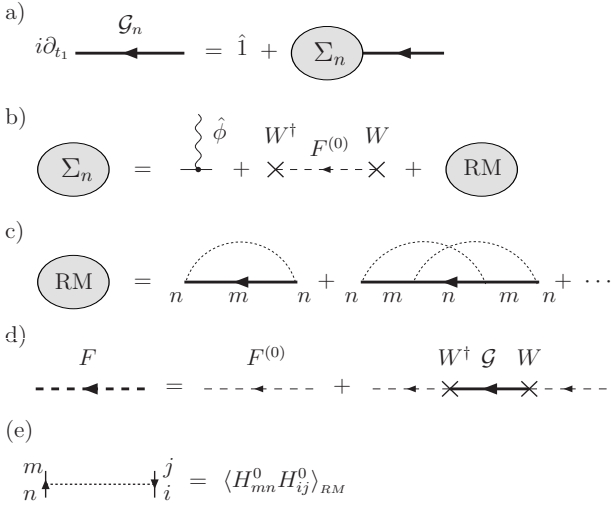}
\caption{Dyson equation for the ensemble averaged dot Green function (a-c), and the expression for the lead Green function (d). 
The crossing diagrams in (c) are smaller than the non-crossing one by powers of $1/M$ and can be ignored. The disordered matrix elements in (e) are given by \req{RMT}.
}
\label{fig5}
\end{figure}

{\setlength{\arraycolsep=0pt}The charge channel of the universal interaction Hamiltonian \req{hint} is given by
\be
H_{int} = E_c N^2,
\ee
where the number operator $N$ was defined in \req{nhat}. This interaction can be decoupled via the Hubbard-Stratonovitch 
transformation by introducing a scalar fluctuating field $\hat{\phi}$ with the 
following matrix structure in the Keldysh space:
\be\label{phik}
\hat{\phi} = \left( \begin{array}{cc }
      \phi_+ & \phi_-   \\
     \phi_- & \phi_+  
\end{array} \right)_K.
\ee
This field modifies the self-energy of the electrons by contributing a term proportional to $\hat{\phi}$  [see \req{predyson} and Fig.~\ref{fig5}.b] .
For the purpose of the one loop approximation we will employ in the next section, these fields
can be considered as Gaussian with the propagators\begin{subequations}
\label{dprop}
\bea
 \langle \phi_+(t_1)\phi_+(t_2)\rangle_\phi& = & \frac{i}{2}\D_\phi^K(t_1,t_2), \\
 \langle \phi_+(t_1)\phi_-(t_2)\rangle_\phi& = &\frac{i}{2}\D_\phi^R(t_1,t_2), \\
 \langle \phi_+(t_2)\phi_-(t_1)\rangle_\phi  & = & \frac{i}{2}\D_\phi^A(t_1,t_2),\\
 \langle \phi_-(t_2)\phi_-(t_1)\rangle_\phi  & = & 0.
\eea\end{subequations}
In the saddle point approximation, the propagators are solutions of the matrix (in Keldysh space) Dyson 
equation 
\bea\label{mdes}
&\!\hat{\D}_\phi = 2E_c \left( \hat{1}+\hat{\Pi}_\phi\hat{\D}_\phi\right),&\\
&\hat{\D}_\phi = \begin{pmatrix}
  \D_\phi^R    &  \D_\phi^K  \\
      0 &  \D_\phi^A
\end{pmatrix}_K, \qquad
& \hat{\Pi}_\phi = \begin{pmatrix}
  \Pi_\phi^R    &  \Pi_\phi^K  \\
      0 &  \Pi_\phi^A
\end{pmatrix}_K, \nonumber
\eea
where $\hat{1}=\hat{1}_K\delta(t_1-t_2)$, and 
\bea\label{polG}
&&\Pi_\phi^R(t_1,t_2) = \Pi_\phi^A(t_2,t_1) \!=\!
\sum_{n=1}^{M}\! \frac{ \delta \Tr_s(\mathcal{G}_{nn}^K(t_1,t_1|\phi)) }{2i\delta \phi^{+}(t_2)},\nl
&&\Pi_\phi^K(t_1,t_2) = \sum_{n=1}^{M}\frac{ \delta \Tr_s(\mathcal{G}_{nn}^K(t_1,t_1|\phi)+\mathcal{G}_{nn}^Z(t_1,t_1|\phi)) }{2i\delta \phi^{-}(t_2)}.\nl
\eea
Here we have introduced the Green function of the dot electrons as a functional of 
the field $\hat{\phi}$ in its matrix form in Keldysh space:
 \be\label{green2}
\hat{\mathcal{G}}_{nm} (t_1,t_2|\phi)= \left( \begin{array}{cc }
      \mathcal{G}^R_{nm}(t_1,t_2|\phi) &  \mathcal{G}^K_{nm}(t_1,t_2|\phi)  \\
     \mathcal{G}^Z_{nm}(t_1,t_2|\phi) & \mathcal{G}^A_{nm}(t_1,t_2|\phi) 
\end{array} \right)_K,
\ee
such that its average over the fluctuating field $\hat{\phi}$ gives the usual 
expressions as in \req{rak}, with the averaged $\mathcal{G}^Z$ vanishing. We will suppress the argument $\phi$ in the subsequent formulas.}

We construct perturbation theory  for the RM averaged Green function $\langle \hat{\mathcal{G}}_{nm}\rangle_{_{RM}}= 
\mathcal{G}_{n} \delta_{nm}$, as an expansion in powers of the random matrix Hamiltonian $H_0$, and the Hubbard-Stratonovitch field $\hat{\phi}$ (see Fig.~\ref{fig5} for the diagrammatic form of the Dyson equation\cite{abrikosov} for $\mathcal{G}_n$).  
The first term in the self energy expansion (Fig.~\ref{fig5}.b) is due to the effect of the Hubbard-Stratonovitch field $\hat{\phi}$, which only changes the phase of the Green function and does not lead to scattering between different orbitals. The second term is due to scattering to and from the leads. The third term  (Fig.~\ref{fig5}.c) gives the contribution of the non-interacting closed dot Hamiltonian $H^0$ after averaging over the random matrix ensemble, where the random matrix correlation in panel (e) is given by  (\ref{RMT}). The crossing diagrams in Fig.~\ref{fig5}.c are smaller than the 
non-crossing diagram by a factor of $1/M$, and will be neglected. In addition, the second term in (\ref{RMT}) contributes to the non-crossing diagram only when $m=n$ (see Fig.~\ref{fig5}.c), and is therefore small by a factor of $1/M$. This means that the orthogonal and unitary ensembles are equivalent for charge 
and triplet channel (the Cooper channel case is considered in 
Sec.~\ref{sec:usacoop}). 
Thus, in the $M\rightarrow \infty$ limit, we obtain the following Dyson equation for $\mathcal{G}_n$
\bea\label{predyson}
i \frac{\partial \hat{\mathcal{G}}_n}{\partial t_1}& = &\one + \hat{\Sigma}_n \hat{\mathcal{G}}_n, \\
\hat{\Sigma}_n&=&  \hat{\phi}  -i \pi \nu W_{n \alpha} \hat{f}_\alpha W^\dagger_{\alpha n}  + M \left(\frac{\delta_1}{\pi}\right)^2 \sum_{m=1}^{M}  
\hat{\mathcal{G}}_m.\nonumber
\eea
Here, as in subsection \ref{sec:leads} the products are understood as operator (matrix) multiplication in time as well as in Keldysh and spin spaces. In particular for the unit operator we have 
$\one=\delta(t_1-t_2)\one_K\otimes 1_s$, and $\hat{\phi}$ is understood as the operator $\hat{\phi}(t_1)\delta(t_1-t_2)1_s$. We have also defined
\be\label{deff}
\hat{f}_\alpha\equiv \frac{i}{\pi\nu} 
\int \frac{d k}{2\pi} e^{-\eta |k|}  \hat{F}^{(0)}_\alpha (k),
\ee
where $\hat{F}^{(0)}_\alpha (k)$ was defined in \req{f0}. Using \req{f0} we integrate over $k$ obtaining
\be
\hat{f}_\alpha(t_1-t_2) = \int \frac{d \epsilon}{2\pi} e^{-i \epsilon( t_1 - t_2)} \hat{f}_\alpha(\epsilon),
\ee
where
\be\label{fhatform}
\hat{f}_\alpha(\epsilon) =  \left( \begin{array}{cc}
1 & f^K_\alpha(\epsilon) \\ 0 & -1
\end{array}\right)_K\otimes 1_s,
\ee
and the function $f^K_\alpha(\epsilon)$ was defined in \req{fkeq}. 
Substituting \req{dubya} into \req{predyson} and defining 
\bea\label{deffg}
\hat{g}_n \equiv  i M \,\frac{\delta_1}{\pi}\, \hat{\mathcal{G}}_n, &\qquad&
\hat{g}\equiv \sum_{n=1}^{M}\, \frac{\hat{g}_n}{M},
\eea
we obtain
\be\label{dyson}
\frac{1}{M}\frac{\partial \hat{g}_n}{\partial t_1} + i \frac{1}{M} \hat{\phi}\,  \hat{g}_n  + 
\tn\fhat  \hat{g}_n +   \hat{g} 
\hat{g}_n  = \one,
\ee
where we have adopted units such that $\delta_1/\pi=1$:
\[ t\rightarrow \frac{t}{\delta_1/\pi} \hspace{1 cm} \phi\rightarrow\ \phi\ \frac{\delta_1}{\pi} . \]

As we are interested in the limit $M \to \infty$ of the RMT, we can neglect the
first two terms on the left side of \req{dyson}, obtaining
\bea \label{constr}
\tn\fhat  \hat{g}_n +   \Ghat  \hat{g}_n  = \one, && (n\leq \nc),\nl
\Ghat  \hat{g}_n  = \one, && (n > \nc).
\eea
Summing \req{constr} over all $n$ and neglecting terms of order $\nc/M$ we obtain the following constraint for $\hat{g}$:
\be\label{eq:cons}
\Ghat \cdot \Ghat = \one.
\end{equation}
Using \reqs{constr}--(\ref{eq:cons}) we solve for $\ghat$'s, obtaining
\be\begin{split}\label{gn}
\ghat & =\frac{\one}{\Ghat+\tn\fhat}=
\Ghat(\one+\tn\fhat \Ghat)^{-1}\\ & =(\one+\tn\Ghat  \fhat )^{-1}\Ghat.
\end{split}\ee
Thus, given $\hat{g}$ and the lead Green functions \req{fhatform}, we can completely determine the $\hat{g}_n$'s using \req{gn}. As for $\hat{g}$,  \req{eq:cons} in general only constrains $\hat{g}$ to a certain manifold, and does not determine it further. 
The non-equilibrium evolution of $\hat{g}$ on this manifold is given by the terms of order $1/M$ in \req{dyson} which were neglected in reducing that equation to  \req{constr}. In order to seperate those $1/M$ terms, we subtract from \req{dyson} its transpose, obtaining
\be \begin{split}\label{1}
\frac{1}{M}\frac{\partial\hat{g}_n}{\partial t} + i\frac{1}{M}\,[\hat{\phi}\,;\hat{g}_n] = 
 - \tn [\, \hat{f}_n\, ;\hat{g}_n] + [\, \hat{g}_n\, ; \hat{g}] \, ,\nonumber
\end{split}\ee
which after summation over all $n$ gives
\be\label{eil}
\frac{\partial\hat{g}}{\partial t}  + i\,[\, \hat{\phi}\, ;\hat{g}] = - \sum_{n=1}^{\nc}\;\tn [\, \hat{f}_n\, ;\hat{g}_n]\, .
\ee
Here $\partial/\partial t \equiv \partial/\partial t_1 +
\partial/\partial t_2$ is the derivative with respect to the ``center
of mass'' time, and $[a;b] = ab-ba$ is the commutator. 
Using \req{constr}, we can rewrite \req{eil} as
\bea\label{usadel}
\frac{\partial\hat{g}}{\partial t} +  i\,[\, \hat{\phi}\, ;\hat{g}] = \sum_{n=1}^{\nc}\;[\hat{g}\, ;\hat{g}_n],
\eea
which we call the 0D Usadel equation.\cite{usadel} 
It is easy to see that this equation is consistent with the constraint \req{eq:cons}. 

We are now prepared to take the $M \to \infty$ limit in  the formula for the current \req{tofu}. 
Using the definition \req{deffg} and constraint \req{constr} in \req{tofu} we obtain
\bea\label{apcur}
I_a & =&  \frac{e}{2} \sum_{n\in a}  \Tr_s\left(-\,\tn[\, \fhat\, ;\ghat]^K\right),\nl
&=& \frac{e}{2} \sum_{n\in a}   \mathrm{Tr}_s [\Ghat\, ;\ghat]^K.
\eea
Thus the total outgoing current is proportional to the Keldysh part of the right hand side of the Usadel equation, \req{usadel}, which states nothing but the conservation of number of particles.

We can rewrite the right hand side of \req{usadel} in a form that is more convenient for later use. Using \req{gn}, we write this commutator in terms of $\Ghat$ and
$\fhat$ only:
\begin{eqnarray}\label{s2}
[\Ghat\, ;\ghat] & = & (\one+\tn\fhat  \Ghat)^{-1} - (\one+\tn\Ghat  \fhat )^{-1}\nonumber\\
& = & \tn [\, \Ghat\, ;\fhat]\cdot \ghat \cdot \ghat\\
& = & \frac{1}{4} T_n [\, \Ghat\, ;\fhat] \cdot 
\left[\one + \frac{1}{4}T_n\left(\big\{\Ghat;\fhat\big\} - \hat{2}\right)\right]^{-1},
\nonumber
\end{eqnarray}
where $\{a;b\}= ab + ba$ is the anti-commutator, and in obtaining the third line from the second, we have used  \req{gn} and \req{eq:cons} to write 
\bea\label{gngn}
\ghat \cdot \ghat & = &[(\one+\tn\fhat \Ghat)\cdot(\one+\tn\Ghat
\fhat )]^{-1}
\nl &=&\left[(1+\tn)^2\one + \tn \left(\big\{\Ghat;\fhat\big\} -
    \hat{2}\right)
\right]^{-1}\nonumber.
\eea
Here the  transmission coefficient of the $n$-th channel is given by
\begin{equation}\label{ttnn}
T_n \equiv\ \frac{4 \tn}{(1+\tn)^2}.
\end{equation}

To obtain the polarization operators, we average \req{polG} over realizations of the RM. \footnote{Ensemble averaging of the polarization operator separately from the Green function is allowed because of the large conductance $G_0\gg G_q$.} Using \req{deffg}, we obtain
\bea\label{polr}
\Pi_\phi^R(t_1,t_2)& = &-\frac{2}{\pi}\delta(t_1-t_2) 
- \frac{ \delta \Tr_s(g^K(t_1,t_1)) }{2\delta \phi^{+}(t_2)},\nl
\Pi_\phi^K(t_1,t_2)&=& - \frac{ \delta \Tr_s(g^K(t_1,t_1)+g^Z(t_1,t_1)) }{2\delta \phi^{-}(t_2)}.
\eea
The appearance of the $\delta$-function term here deserves comment. It is analogous to the usual ultraviolet anomaly.  Namely,  
in the first line of \req{polr} the expression $g^K(t_1,t_1)$ is understood as the limit $t_1\to t_1'$ of $g^K(t_1,t_1')$, where the latter is calculated in the $M\to\infty$ limit already. On the other hand, \req{polG} implies the opposite order of limits, and the $\delta$-function takes care of this discrepancy:
\bea
&& -\frac{2}{\pi}\delta(t_1-t_2) =  \\ && \ \left[\lim_{M\to\infty}\lim_{t_1\to t_1'} - \lim_{t_1\to t_1'}\lim_{M\to\infty}\right] 
 \sum_{n=1}^{M}\! \frac{ \delta \Tr_s(\mathcal{G}_{nn}^K(t_1,t_1'|\phi)) }{2i\delta \phi^{+}(t_2)}. 
\nonumber
\eea
This equality can be checked by explicit calculation. In fact, the coefficient in front of the delta function is guarded by the requirement of gauge invariance, i.e. the total electron number response to the potential $\phi_\omega$ should vanish for $\omega \gg 1/\tau$.

Equation (\ref{usadel}) together with the constraint (\ref{eq:cons}), the
identity (\ref{s2}), the propagators \req{mdes} and the polarization operators \req{polr}, completely determine the kinetics of the quantum dot, given the distribution functions of the electrons in the leads. 
Moreover, the last equality in \req{s2} gives a convenient starting point for 
the perturbation theory; indeed
in the non-interacting case, $\Ghat$ has the same structure as $\fhat$,
\req{fhatform}, and the square bracket in the last line of \req{s2} 
reduces to the identity (and the commutator gives the relaxation of $\Ghat$ 
towards $\fhat$).

\subsection{Triplet Channel} 
\label{usadel:triplet}
The triplet part of the interaction \req{hint} is given by
\be
H_{int} = J_s \vec{S}^2,
\ee
where the total spin operator $\vec{S}$ was defined in \req{spin}. This can similarly be decoupled using a vector Hubbard-Stratonovitch field $\vec{\hat{h}}$, with each component having the same Keldysh structure as in (\ref{phik}). The presence of this field leads to the replacement $\hat{\phi} \to \hat{\phi} + \hat{h}$,
\be\label{hstruc}
\hat{h}\equiv \sum_i \hat{h}_i \otimes \sigma^i,
\ee
 in the self-energy \req{predyson}. Here $i,j=x,y,z$, and $\sigma^i$ are the Pauli matrices in spin space. 
The propagators for this field are defined as in \req{dprop} by replacing the $\phi$'s with different components of  the vector field $\vec{h}$, such that now each Keldysh component of the propagator acquires a $3\times 3$ tensor structure. In the saddle point approximation we have 
\be\label{tdes}
\!\hat{\D}_h =  \frac{J_s}{2} \left( \hat{1}+\hat{\Pi}_h\hat{\D}_h\right),
\ee
where $\one= \hat{1}_K \delta_{ij}\delta(t_1-t_2)$, and the polarization operator tensors are given similarly as in \req{polG} by variational derivative of the Green functions with respect to components of the field $\vec{\hat{h}}$. Repeating the steps leading to \req{polr}, we obtain
\bea\label{polrt}
\left[\Pi_h^R\right]_{ij}(t_1,t_2)& = &-\frac{2\delta_{ij}}{\pi}\delta(t_1-t_2) 
- \frac{ \delta \Tr_s(\sigma^i g^K(t_1,t_1)) }{2\delta h_j^{+}(t_2)}\nl
\left[\Pi_h^K\right]_{ij}(t_1,t_2)&=& - \frac{ \delta \Tr_s(\sigma^i(g^K+g^Z)(t_1,t_1)) }{2\delta h_j^{-}(t_2)},
\eea
where $i,j=x,y,z$. 
Furthermore, in the presence of a magnetic field in the $z$-direction, we must add to the 
dot's Hamiltonian the Zeeman energy term
\bea\label{zeemanham}
H_Z = E_Z \hat{z}\cdot\vec{S}, && (E_Z = g_L\mu_B B),
\eea
where $g_L$ is the Lande g-factor and $\mu_B$ is the Bohr magneton.
This causes the $z$-component of the field $\vec{h}$ to acquire a non-zero average at the saddle point, given by
\be\label{haverage}
\langle h^+_z \rangle =  \int dt_2 [\D^R_h \Pi^R_h]_{zz}(t_1,t_2)\frac{E_Z}{2},
\ee
where the other components of the tensor vanish due to the symmetry of spin rotations around the $z$-axis. Here, the factor $1/2$ arises from the same factor accompanying the Pauli matrix in the definition of the operator $\vec{S}$, \req{spin}.  
Separating this average from the fluctuating part, and redefining $\vec{h}$ to stand for the latter, we obtain the modified Usadel equation:
\bea\label{tripusadel}
\frac{\partial\hat{g}}{\partial t} +\frac{1}{2} i E_Z^*\,[\,\sigma_z\, ; \hat{g}]+  i[ \hat{h}\, ;\hat{g}] = \sum_{n=1}^{\nc}\;[\hat{g}\, ;\hat{g}_n],
\eea
where $E_Z^*$ is the renormalized Zeeman energy
\be\label{ezrenor}
E_Z^* \equiv E_Z + 2\langle h^+_z \rangle.
\ee

Equation (\ref{tripusadel}) together with  \reqs{tdes}--(\ref{polrt}), and the constraint (\ref{eq:cons}) and identity \req{s2} which are still valid, completely determine the kinetics of the quantum dot under the effect of spin fluctuations. 

\subsection{Cooper Channel}\label{sec:usacoop}
Interaction in the Cooper channel is given by
\be
H_{\mathrm{int}} = J_c \mathcal{T}^{\dagger}\mathcal{T},
\ee
where the pairing operator $\mathcal{T}$ was defined in \req{T}.
This interaction can be decoupled 
using a complex Hubbard-Stratonovitch field $\hat{\Delta}$ 
which also has the structure (\ref{phik}) in Keldysh space. This pairing field gives rise to anomalous Green functions which can be taken into account by introducing the standard Gor'kov-Nambu\cite{gorkov:nambu} (GN) spinor 
\be
\Psi \equiv \left(\! \begin{array}{c}
      \psi   \\
      i \sigma^y \psi^\dagger  
\end{array}\!\right),
\ee
where $\sigma^y$ acts in the spin space, and we denote Pauli matrices in the GN space by $\tau^i$. We also redefine the dot and lead Green functions $ \hat{\boldsymbol{\mathcal{G}}}_{nm}$ and $\hat{\boldsymbol{F}}_{\alpha \beta}$  to be $2\times 2$ matrices in the GN, as well as in the Keldysh and spin spaces (we will use bold symbols to represent matrices in GN space), such that after averaging over the Hubbard-Stratonovitch field different Keldysh components will have the same structure as in \req{rak}, with $\psi$ and $\psi^\dagger$ replaced by $\tau^z \Psi$ and $\Psi^\dagger$ respectively, and the products understood as direct products in GN as well as in spin spaces. Because of the standard 
 convention to include $\tau^z$ in the definition of GN Green functions, the time derivatives in the properly modified \reqs{predyson} and (\ref{usadel}) will be accompanied by this Pauli matrix [see 
(\ref{usacooper}) below].  We introduce the notation
\be\label{namstruc}
\hat{\boldsymbol{g}} = \begin{pmatrix}
      \hat{g} & \hat{\ff}   \\
      \hat{\bar{\ff}} & \hat{\tilde{g}} 
\end{pmatrix}_N
\ee
to represent the GN components of the Green function $\hat{\boldsymbol{g}}$ (related to $\hat{\boldsymbol{\mathcal{G}}}_n$'s as in \req{deffg}). The upper left GN component of this matrix gives the Green function used in the singlet and triplet channels, and the off-diagonal components are the anomalous Green functions. Since the components of the GN spinor are not independent ($\Psi = \sigma^y\otimes\tau^y \Psi^\dagger$), the components of the Green function $\hat{g}$ are not entirely independent either, and are related by
\bea\label{grelation}
\hat{\boldsymbol{g}} = \hat{\boldsymbol{\mathrm{S}}}\,\, \hat{\boldsymbol{g}}^{\mathrm{T}}\, \hat{\boldsymbol{\mathrm{S}}}, &\qquad& (\hat{\boldsymbol{\mathrm{S}}} \equiv \sigma_{K}^x \otimes \sigma^y \otimes \tau^x),
\eea
where $\sigma_{K}^x$ is the Pauli matrix acting in the Keldysh space, and the superscript $\mathrm{T}$ stands for transposition of the operator $\hat{g}$ in time, Keldysh, spin and GN spaces. The same relation is satisfied by $\hat{\boldsymbol{f}}_n$'s so that their explicit form is 
 \be
\hat{\boldsymbol{f}}_n = \begin{pmatrix}
     \hat{f}_n &   0 \\
     0 & \sigma^x_K \hat{f}^{\mathrm{T}}_n\sigma^x_K  
\end{pmatrix}_N,
\ee
and the matrix $\hat{f}_n$'s are given by \req{fhatform}. 
We define the matrix $\hat{\boldsymbol{\Delta}}$ and the one-particle Hamiltonian $\boldsymbol{H}^0$ to have the following forms in GN space:
\bea\label{hcoop}
\hat{\boldsymbol{\Delta}} \equiv \begin{pmatrix}
     0 & \Delta   \\
      -\Delta ^*& 0 
\end{pmatrix}_N , &&
\boldsymbol{H}^0 \equiv \begin{pmatrix}
     H^0 & 0  \\
      0 & \sigma^y H^{0T} \sigma^y
\end{pmatrix}_N,
\eea
where we have included the Zeeman splitting Hamiltonian \req{zeemanham} in $H^0$.
With these definitions the expression for self-energy (\ref{predyson}) takes the form (in units where $\delta_1/\pi = 1$)
\be\label{selfenergycoop}
\hat{\boldsymbol{\Sigma}}_n = \hat{\boldsymbol{\Delta}} + \frac{\boldsymbol{E}_Z^*}{2}- i\pi\nu W_{n\alpha}\hat{\boldsymbol{f}}_\alpha W^\dagger_{\alpha n} -i M \hat{\boldsymbol{g}} + i \frac{\rmg_h}{4}\hat{\boldsymbol{\ff}},
\ee
where we have defined 
\be
\boldsymbol{E}_Z^*\equiv E_Z^*\sigma^z\otimes\tau^z\otimes\one_K,
\ee
and
\be
\hat{\boldsymbol{\ff}}\equiv
\begin{pmatrix}
      0 & \hat{\ff}   \\
      \hat{\bar{\ff}} & 0 
\end{pmatrix}_N.
\ee
The origin of the last term in \req{selfenergycoop} is explained as follows: in the presence of a time reversal symmetry breaking 
magnetic field, $H^0$ and $H^{0\mathrm{T}}$ in \req{hcoop} are not equal. Therefore, to leading order in $1/M$, the non-crossing diagram of Fig.~\ref{fig5}.c (being proportional to $\langle\hat{H}^0\hat{g}\hat{H}^0\rangle_{_{RM}}$) receives contributions from the 
second term in the right hand side of \req{RMT} in the off-diagonal GN parts. 

Using \req{selfenergycoop} and following steps similar to the ones leading to \req{usadel}, we arrive at the properly modified 
Usadel equation for the 
GN Green functions
\be\begin{split}\label{usacooper}
\tau^z \frac{\partial\hat{\boldsymbol{g}}}{\partial t_1} + \frac{\partial\hat{\boldsymbol{g}}}{\partial t_2} \tau^z + \frac{i}{2}[ \boldsymbol{E}_Z^*\, ;\hat{\boldsymbol{g}}]  - \frac{\rmg_h}{4}[\hat{\boldsymbol{\ff}};\hat{\boldsymbol{g}}]= \\
 - i[ \hat{\boldsymbol{\Delta}}\, ;\hat{\boldsymbol{g}}] + \sum_{n=1}^{\nc}[\hat{\boldsymbol{g}}\, ;\hat{\boldsymbol{g}}_n].
 \end{split}\ee
We note that in order to take into account the interaction in all three channels together, we can simply add the fields $\hat{\boldsymbol{\phi}}=\hat{\phi}\otimes 1_N$ and $\hat{\boldsymbol{h}}=\hat{h}\otimes\tau^z$ to $\hat{\boldsymbol{\Delta}}$ in \req{usacooper}.  


For repulsive interaction ($J_c>0$) or for attractive interaction  ($J_c<0$) at $T>T_c$, the field $\hat{\Delta}$ can be considered as Gaussian with propagators 
\bea\label{cprop}
  \langle \Delta_+(1)\Delta_+^*(2)\rangle_{_\Delta}  & = & \frac{i}{2}\D_\Delta^K(1,2),\\
 \langle \Delta_+(1)\Delta_-^*(2)\rangle_{_\Delta} & = & \frac{i}{2}\D_\Delta^R(1,2), \nonumber\\
 \langle\Delta_-(1) \Delta_+^*(2)\rangle_{_\Delta} & = & \frac{i}{2}\D_\Delta^A(1,2), \nonumber\\
 \langle\Delta_-(1) \Delta_-^*(2)\rangle_{_\Delta} & = & 0.\nonumber
\eea
In the saddle point approximation, the propagators satisfy the matrix (in Keldysh space) Dyson equation 
\be\label{mdescoop} 
\hat{\D}_{\Delta} = J_c \left(\one + \hat{\Pi}_{\Delta}\hat{\D}_{\Delta}\right),
\ee
where $\one=\one_K\delta(t_1-t_2)$, and the polarization operators are given by
\bea\label{polcoop}
\Pi_\Delta^R(t_1,t_2) &=& \Pi_\Delta^A(t_2,t_1) 
= \frac{1}{4} \frac{\delta \mathrm{Tr}_s(\ff^K(t_1,t_1))}{\delta \Delta_+(t_2)}
\, , \nl
\Pi_\Delta^K(t_1,t_2) & =&
\frac{1}{4} \frac{\delta\mathrm{Tr}_s( \ff^K(t_1,t_1)
)}{\delta \Delta_-(t_2)}.
\eea

Equation (\ref{usacooper}) together with  \reqs{mdescoop}--(\ref{polcoop}), and the constraint (\ref{eq:cons}) and identity \req{s2} (which are still valid with the additional GN structure understood), constitute a complete description of the kinetics of the quantum dot under the effect of pairing fluctuations.

\section{Derivation of the Kinetic Equation}\label{sec:kineq}

Even though the Usadel equation (\ref{usacooper}) gives a complete description of the kinetics of the open quantum dot, its solution for a general form of the Hubbard-Stratonovitch fields is not tractable due to its non-linearity and non-locality in the time domain.
The purpose of this section is to reduce the Usadel equation to the kinetic equation for all three channels of the interaction. The kinetic equation describes processes characterized by the time $\Delta t$ which are slow. In the particular case of the quantum dots, slowness means $\Delta t \gg 1/\omega^*$, where $\omega^*$ is the characteristic energy transfer in the electron-electron interaction, (the scale $\omega^*$ is determined by the shape of the distribution function $f^K(\epsilon)$, e.g. for thermal equilibrium $\omega^*\simeq T$) . The relation between $\Delta t$ and the relaxation scale determined by the kinetic equation itself (i.e. $1/\tau$ defined in \req{gma0}) may be arbitrary (see e.g. Sec.~II B of Ref.~\onlinecite{catelani:aleiner:05} for more detailed discussion).

 In the first loop approximation for the interactions, the different channels are decoupled and we will treat them seperately. 
 
  \subsection{Charge Channel}\label{sec:cc}

\subsubsection{Gauge Transformation}\label{subsec:gt}

The condition that justifies the first loop approximation in spite of the large charging energy ($E_c \gg \delta_1$), is the largeness of the leads' conductance which suppresses the effect of fluctuations of the Hubbard-Stratonovitch field for small frequency $\omega \simeq 1/\tau \gg \delta_1$.
On the other hand, if the transmitted frequency is larger than $1/\tau$, the dot can be considered as closed. For the closed dot, the field $\hat{\phi}$ is coupled only with the total number of electrons, $N$, which commutes with the Hamiltonian. This interaction causes the motion of the one-particle energy levels without any redistribution of particles between them, which even though produces a large, singular effect in the one-particle Green function (known as Coulomb blockade), has nothing to do with relaxation inside the dot. To eliminate the effect of the trivial motion of the levels from the very beginning, we invoke a gauge transformation (first proposed in Ref.~\onlinecite{kamenev:andreev:99}) in the Usadel equation (\ref{usadel}):
\be\label{gts}
\hat{g}(t_1,t_2) = e^{-i \hat{K}(t_1)} \hat{\tilde{g}}(t_1,t_2) e^{i \hat{K}(t_2)},
 \ee 
 where the matrix field 
 \be
 \hat{K} = \begin{pmatrix}
      K_+&K_-    \\
      K_-&  K_+
\end{pmatrix}_K,
\ee 
will be chosen such that almost all of the contributions of $\hat{\phi}$ from frequencies $\omega \gtrsim 1/\tau$, are eliminated. \cite{kamenev:andreev:99,catelani:aleiner:05} The corrections to $\hat{\tilde{g}}$ can then be found by perturbation theory. 

We substitute \reqs{s2} and (\ref{gts}) into \req{usadel} and obtain
\bea\label{gtusadel}
&&\frac{\partial\hat{\tilde{g}}}{\partial t} +  i\,[\, \hat{\phi}\, -\partial_t \hat{K};\hat{\tilde{g}}] =  \\ 
&&\frac{T_n}{4}  [\, \hat{\tilde{g}}\, ;e^{i\hat{K}}\fhat e^{-i\hat{K}}]\! \cdot\!
\left[\one + \frac{T_n}{4}\left(\big\{\hat{\tilde{g}}; e^{i\hat{K}}\fhat e^{-i\hat{K}}\big\} \! - \hat{2}\right)\right]^{-1}\!.\nonumber
\eea
 We will look for a perturbative solution to \req{gtusadel} in the first loop approximation; to do so it suffices to retain terms that are at most quadratic in the field $\hat{K}$ on the right hand side. At zeroth order in $\hat{\phi}$, the Green function has the form
\be\label{gform}
 \hat{\tilde{g}}=  \left( \begin{array}{cc}
\delta (t_1-t_2) & \gk (t_1,t_2) \\ 0 & -\delta (t_1-t_2)
\end{array}\right)_K.
\ee
We will require that this form is preserved even in the first and second orders, i.e. the corrections to the spectrum (given by $g^R$ and $g^A$) are indeed eliminated by the gauge transformation. Considering the $R, A$ and $Z$ components of \req{gtusadel} in linear order in $\hat{K}$ ,  we see that this condition is satisfied if  
\be\label{phinegkneg}
\left(\frac{\partial}{\partial t} -\glr\right) K_- (t) = \phi_-(t),
\ee 
so that 
\be\label{kmeq}
K_- = - \bar{\cal L}^g \phi_-.
\ee
Here the bar indicates complex conjugation,  and the 
0-dimensional diffuson ${\cal L}^g$ is given by
\bea
{\cal L}^g (t)&=&\int\!\frac{d\omega}{2\pi} \, e^{-i\omega t} \, 
{\cal L}^g (\omega) \label{FT} \nl
\label{diffdef}
{\cal L}^g (\omega)&=&\frac{1}{-i\omega + 1/\tau} \, , 
\eea
with $1/\tau$ the escape rate \req{gma0} in units where $\delta_1=\pi$
\be
 \, \frac{1}{\tau} 
\!=\!\sum_n \frac{T_n}{2}\!=\!\frac{\rmg_L\!+\rmg_R}{2},
\ee 
and we introduced the operator notation
\be\label{opnot}
\left[\bar{\cal L}^g \phi_- \right] (t_1) = \int\!dt_3 \,
\bar{\cal L}^g (t_1 -t_3) \phi_- (t_3) \, .
\ee
The diffuson ${\cal L}^g(t)$ gives the classical probability for an electron introduced to the dot at $t=0$ to remain in it at time $t$. Note that in the present case the diffuson is much simpler than in higher 
dimensional systems, where it depends also on position in space and direction 
of momentum on the Fermi surface. The same is true for the fields 
$\hat{\phi}$ and $\hat{K}$.

To solve the Keldysh component of the gauge transformed Usadel equation (\ref{gtusadel}),
we will look for $K_+$ in the form
\be\label{kpeq}
\left[{\cal L}^g\right]^{-1} K_+ = \phi_+ - 2\tilde{K}_- \: , \quad
\tilde{K}_- \equiv (i\partial_{t})^{-1} M K_-,
\ee
where the operator $M(t_1,t_2)$, will be chosen to 
simplify further expansion of $\gk$. Substituting \req{kpeq} into the Keldysh component of \req{gtusadel}, and expanding the right hand side to second order in $\hat{K}$ we obtain:
\bea \label{gkusadel}
&&\frac{\partial g^K}{\partial t} = \sum_n \frac{T_n}{2} (f^K_n - g^K) \\
\nonumber
&& + i \sum_n \frac{T_n}{2} \left[K_+ ; f^K_n - g^K\right] + i Q K_-  - 2i \left[\tilde{K}_- ; g^K\right] \\
\nonumber
&&- \sum_n \frac{ T_n}{4}\Big\{ \! \left[K_+;\left[K_+; f^K_n\right]\right] \\
\nonumber
&& \qquad \qquad + (1- T_n) \left[ K_+ f^K_n K_- g^K  -  g^K K_- f^K_n K_+  \right] \\[3pt]
\nonumber
&&\qquad \qquad + T_n\left[  K_+ f^K_n K_- f^K_n -  f^K_n K_- f^K_n  K_+\right] \!\Big\}.
\eea
Here we do not display second order terms that vanish after averaging over the field $\hat{\phi}$, e.g. $K_- K_-$ (see \req{kprop}) and $K_+(t_3)K_-(t_3)$. The latter product vanishes due to the analytic properties of the retarded propagator (see \reqs{kprop} and (\ref{KR})). Defining $\delta \gk$ to be the linear order correction to $g^K$,
we find:
\begin{subequations}
\bea
&& \delta \gk  = \delta \gk_+ + \delta \gk_-,\nl
&&\left( \frac{\partial}{\partial t} + \frac{1}{\tau} \right) \delta \gk_+ = i \sum_n \frac{T_n}{2}
\left[ K_+ ; f_n^K - g^K  \right] ,\qquad \label{dgkpeq} \\
&& \left( \frac{\partial}{\partial t} + \frac{1}{\tau}\right) \delta\gk_- = i 
Q K_- -2 i \left[ \tilde{K}_- ; g^K \right], \label{dgkneg}
\eea\end{subequations}
where 
the $K$'s are understood as operators: $K(t_1,t_2) = K(t_1)\delta (t_1-t_2)$.
The operator $Q(t_1,t_2;t_3)$ acts on $K_-$ as
\be
\left[Q K_-\right] (t_1,t_2) = \int\!dt_3 \, Q(t_1,t_2;t_3) K_-(t_3),
\ee 
and is related to products of the Keldysh Green function of the dot and the
leads:
\bea\label{qdef}
&&Q (t_1,t_2;t_3) = \\[3pt]
&& \frac{1}{4}\sum_n \Big\{ T_n 
\left[ f_n^K(t_1,t_3)  f_n^K(t_3,t_2)  +  \gk(t_1,t_3) \gk(t_3,t_2) \right] \nl
&& + T_n ( 1\!- T_n) \left[ (g^K-f_n^K)(t_1,t_3) (f_n^K-g^K)(t_3,t_2) \right]
\!\Big\} \, . \nonumber
\eea
The Keldysh Green functions are singular at coinciding times: 
\be\label{singulargk}
\gk(t_1,t_2) |_{t_2 \to t_1} = -\frac{2i}{\pi(t_1 - t_2)} + \mathrm{regular},
\ee
(similarly for $f_n^K$) and therefore the products in the second line of \req{qdef} are understood as  
\bes
& \gk(t_1,t_3) \gk(t_3,t_2) \equiv \\
&\frac{1}{2}\sum_{\sigma=\pm 1}  \gk(t_1,t_3+\sigma i 0) \gk(t_3+\sigma i 0,t_2),
\end{split}
\ee
and similarly for $f_n^K\!\cdot\! f_n^K$.
We will see later that the operator $Q$ has the meaning of the fluctuations of the charge going into and out of the dot. The third line in \req{qdef} vanishes in equilibrium and represents the non-equilibrium shot-noise, whereas the second line is present even in equilibrium and has the meaning of Nyquist noise. 

As the right hand side of \req{dgkpeq} vanishes for $t_1=t_2$, we see that $\delta \gk_+ (t_1,t_1) = 0$. This means that $\delta \gk_+$ does not contribute to physical quantities of the dot itself (e.g. total charge or total energy in the dot). Furthermore, taking the limit $t_1 \to t_2$ in the right hand side of \req{dgkneg} and using \req{singulargk}, it is seen that the same condition can be imposed on $\delta \gk_-$ if we choose the operator $M$ to be
\be\label{mq}
M(t_1,t_2) = -\frac{\pi}{4} Q(t_1,t_1;t_2) \, .
\ee
We note how the structure of the $Q$ and $M$ operators is similar to that of Eq.~(5.30) of Ref.~\onlinecite{catelani:aleiner:05} for the case of 
disordered metals, the difference being that in that reference the Green functions are characterized by different directions  instead of different channels.  

{\setlength{\arraycolsep=0pt}The next step is to average \req{gkusadel} over the Hubbard-Stratonovitch fields. For this we need the propagators for the fields $K_{\pm}$ defined as
\bea
\langle K_+(t_1) K_+(t_2)\rangle_\phi & = & \frac{i}{2}K^K(t_1,t_2), \nonumber\\
\langle K_+(t_1) K_-(t_2)\rangle_\phi & = & \frac{i}{2}K^R(t_1,t_2), \nonumber\\
\langle K_-(t_1) K_+(t_2)\rangle_\phi & = & \frac{i}{2}K^A(t_1,t_2), \nonumber\\[3pt]
\langle K_-(t_1) K_-(t_2)\rangle_\phi & = & 0. \label{kprop}
\eea
Their relations to the propagators (\ref{dprop}) and their explicit expressions
will be given in subsection \ref{sec:prop}. 

After averaging over the fluctuating fields and dropping from now on the superscript $K$ in the averaged Green functions ($g=\langle g^K\rangle_\phi$), \req{gkusadel} takes the form 
\be\label{dkineq}
\frac{\partial g}{\partial t}  = \sum_n \frac{T_n}{2} \left( f_n - g \right)
 + \St_1 + \St_2,
\ee
where the collision integrals $\St_1$ and $\St_2$ arise from averaging terms of first and second order in $K_\pm$ respectively. To average the second order terms (given by the last three lines of \req{gkusadel}), we replace $g^K$ with its average $g$ -- keeping the deviation from average will result in higher order terms beyond our approximation-- and average the $K_\pm$ pairs using the propagators (\ref{kprop}), obtaining
\begin{subequations}\label{st1}
\bea
&& \St_2 = \St^{in} + \St^{el}, \\
&& \St^{in} (t_1,t_2) = \frac{i}{8}\sum_n  \Big[ T_n \tilde{K}^K (t_1,t_2)
f_n (t_1,t_2)  \label{stinch} \\ &&  - \frac{1}{2}T_n \left( 1-T_n\right) 
\!\int\! d t_3 \left( K^R_{1,3} - K^A_{3,2} \right)  \left(f^n_{1,3}\, g^{\phantom{n}}_{3,2} + g^{\phantom{n}}_{1,3} \, f^n_{3,2} \right)
\ \nonumber \\ &&  -
T_n^2 
\int\!dt_3 \left( K^R_{1,3} - K^A_{3,2} \right)
f^n_{1,3} \, f^n_{3,2}  \Big], \nonumber \\
&& \St^{el} (t_1,t_2) = -\frac{i}{16}\sum_n T_n (1-T_n) \label{stelch} \\
&&  \times \int\!dt_3 \left( f^n_{1,3}\, g^{\phantom{n}}_{3,2} - g^{\phantom{n}}_{1,3}\, f^n_{3,2}  \right) 
 \left( K^R_{1,3} + K^A_{3,2} \right) , \nonumber
\eea\end{subequations}
where 
\be
\tilde{K}^K (t_1,t_2) = 2 K^K(t_1,t_2) - K^K (t_1,t_1) - K^K(t_2,t_2) \, .
\ee
As for the first order terms (given by the second line of \req{gkusadel}), we write $g^K= g + \delta g^K$, where $\delta g^K$ given by \reqs{dgkpeq}--(\ref{dgkneg}) is the correction linear in $K_\pm$. Keeping terms of second order in $K_\pm$, we obtain
\begin{subequations}\label{st2}
\bea
\St_{1 \: } &=& \langle \St_-\rangle_\phi + \langle \St_+\rangle_\phi 
\, , \\[3pt]
\St_- &=& i \left[\delta Q\right]K_- - 2i\left[\tilde{K}_-; \delta\gk\right], 
\label{stmdef} \\
\St_+ &=& - \frac{i}{\tau} \left[ K_+ ;\delta\gk \right], \label{stp}
\eea
where $\delta Q$ is the first variation of $Q$ [\req{qdef}] with respect to $g^K$:
\be
\delta Q = \frac{\delta Q}{\delta g^K} \cdot \delta g^K_+.
\ee   
Using the solutions of \reqs{dgkpeq}--(\ref{dgkneg}) for $\delta g^K$, one can explicitly average $\St_\pm$ using \req{kprop}, but as it will be shown in the next subsection the collision integral $\St_1$ does not contribute to physical quantities and we will not need its explicit form. 
\end{subequations}}

This formally concludes the derivation of the equation for the Keldysh
Green function in the time domain. However we need to calculate
explicitly the propagators introduced in \req{kprop}; this is done in the 
next subsection.

\subsubsection{Propagators and Collective Excitations}
\label{sec:prop}

The propagators defined in \req{kprop} can be expressed in terms of the 
propagators $\D$, \req{dprop}, thanks to the relations between the fields 
$K_{\pm}$ and $\phi_{\pm}$; see \reqs{kmeq} and (\ref{kpeq}). We obtain 
\bea\label{krldl}
K^R &=& - {\cal L}^g \D_\phi^R {\cal L}^g,\\
K^K &=& {\cal L}^g \D_\phi^K \bar{{\cal L}}^g\nl 
&&+ 2 i \left[ {\cal L}^g \partial_t^{-1} M K^A - K^R M \partial_t^{-1} \bar{{\cal L}}^g\right].\label{kkel}
\eea
Therefore we need to evaluate the $\D$ propagators, given by \req{mdes} and \req{polr}. We note that the Green functions appearing in \req{polr} are the ones before the gauge transformation \req{gts}, so that to linear order in $\hat{\phi}$, the $\delta \hat{g}$'s obtained from \reqs{dgkpeq}--(\ref{dgkneg}) have to be modified according to
\bea
\delta \gk &\to& \delta \gk - i[K_+; g^K] - 2K_- \delta(t_1-t_2),\nl
\delta g^Z &\to& 2 K_- \delta(t_1-t_2).
\eea
We also note that, thanks to \req{singulargk}:
\be
\lim_{t_1\to t_2} - i[K_+; g^K] = -\frac{2}{\pi} \partial_t K_+.
\ee
Using this result together with \reqs{kmeq}--(\ref{kpeq}) and the property
\be\label{dg0}
\delta\gk (t_1,t_1) = 0,
\ee
from \req{polr} we find
\be
\Pi_\phi^R(t_1,t_2) = -\frac{2}{\pi}\big[ \delta(t_1-t_2) - \partial_{t_1} 
{\cal L}^g (t_1,t_2)\big] ,
\ee
or after the Fourier transformation (\ref{FT})
\be\label{polft}
\Pi_\phi^R(\omega) = -\frac{2}{\pi}\frac{1/\tau}{-i\omega+1/\tau}.
\ee
The result for the Keldysh component is
\be\label{polkel}
\Pi_\phi^K= -\frac{4 i}{\pi} {\cal L}^g M \bar{{\cal L}}^g.
\ee
Substituting \req{polft} into the retarded component of \req{mdes} 
we obtain
\be
\D_\phi^R(\omega) = 2 E_c \, \frac{- i \omega + \frac{1}{\tau}}
{-i\omega + \left(1+\frac{4 E_c}{\delta_1}\right)\frac{1}{\tau}},
\ee
where we restored dimensionful units (the difference being the mean level
spacing appearing in the denominator instead of $\pi$). Then we can
calculate the $K^R$ propagator in \req{krldl}; it can be written in terms of 
the diffuson (``ghosts'') propagator ${\cal L}^g$ and the following
propagator ${\cal L}^\rho$ for the collective excitations in the charge channel
\be\label{Lc}
{\cal L}^\rho (\omega) = \frac{1}{-i\omega + 
\left(1+\frac{4 E_c}{\delta_1}\right)\frac{1}{\tau}},
\ee
as
\be\label{KR}
K^R(\omega) = \frac{\delta_1}{2} \frac{1}{-i\omega} \left[ {\cal L}^g (\omega)
-\left(1+\frac{4 E_c}{\delta_1}\right) {\cal L}^\rho (\omega)  \right].
\ee

To find $K^K$ we use the Keldysh part of \req{mdes}:
\be
\D_\phi^K = \D_\phi^R \Pi_\phi^K \D_\phi^A,
\ee
together with \reqs{kkel} and (\ref{polkel}) and obtain
\begin{eqnarray}\label{kkcharge}
K^K &=& - i\frac{\delta_1}{2} \frac{1}{\partial_t}
\bigg\{ \Big[{\cal L}^g {\cal N}^g +{\cal N}^g \bar{\cal L}^g\Big]
\\ && -
\left( 1+ \frac{4E_c}{\delta_1} \right) \Big[{\cal L}^\rho {\cal N}^\rho
+ {\cal N}^\rho \bar{\cal L}^\rho \Big] \bigg\} \frac{1}{\partial_t}, \nonumber
\end{eqnarray}
written in the time domain and in the operator notation. Here the bosonic ``density matrices'' ${\cal N}^\alpha$, $\alpha = g,\rho$, are defined through
\bea\label{defnbos}
({\cal L}^g)^{-1} {\cal N}^g +{\cal N}^g (\bar{\cal L}^g)^{-1} &=& 2 M, \nl
({\cal L}^\rho)^{-1} {\cal N}^\rho+ {\cal N}^\rho (\bar{\cal L}^\rho)^{-1} &=& 2(1+F^\rho) M,
\eea
i.e. they are required to satisfy the kinetic equations
\be\label{boskineq}
\frac{\partial}{\partial t} {\cal N}^\alpha = -2 \left( 1+ F^\alpha \right)
\left( \frac{1}{\tau} {\cal N}^\alpha - M \right),
\ee
with
\be
F^g =0 \, , \quad F^\rho = \frac{4E_c}{\delta_1}.
\ee
While in general we will not solve these kinetic equations, their exsistance
is needed to obtain the conservation law for the energy. To convince the reader
of the bosonic nature of the collective excitations, let us briefly consider
the thermodynamic equilibrium $g(\epsilon)=f_n(\epsilon)=2\tanh (\epsilon/2T)$.
In this case, after Fourier transforming \req{mq} we find
\[
M_{eq}(\omega) = \frac{1}{\tau} \, \omega\coth \frac{\omega}{2T},
\]
and introducing the distribution functions $N^\alpha (t,\omega)$ by
\be\label{bosdistfunc}
{\cal N}^\alpha (t_1,t_2)= \int\!\frac{d\omega}{2\pi} \, e^{-i\omega(t_1-t_2)},
\omega \Big[ 
2N^\alpha\left({\textstyle \frac{t_1+t_2}{2}},\omega\right) + 1 \Big] 
\ee 
we arrive at the followig solution of \req{boskineq}:
\be\label{planck}
N^\alpha_{eq} (\omega) = N_P(\omega) \equiv \frac{1}{e^{\omega/T} -1 },
\ee
i.e. the Planck distribution.

\subsubsection{Conservation Laws and Currents}\label{sec:clc}

The validity of the conservation laws in the kinetic equation approach is 
related to certain properties of the collision integral. 
For example in a closed system the charge conservation law follows from the 
vanishing of the collision integral in the limit 
$t_2 \to t_1$.\cite{catelani:aleiner:05}
In the present case however we are dealing with an open 0-dimensional
system and therefore the conservation law for any physical quantity
like the charge $q(t)$ in the dot should have the form:
\be\label{chcon}
\frac{\partial q}{\partial t} + I =0,
\ee
where $I = I_L + I_R$ represents the total charge flux leaving the dot.

To obtain the charge conservation law, we rely on the following 
general properties of the collision integrals (\ref{st1})-(\ref{st2}):
\be\label{genprop}\begin{split}
& \lim_{t_{2}\to t_{1}} \St^{in} = 0, \\
& \lim_{t_{2}\to t_{1}} \St_+ = 0.
\end{split}\ee
The last property is a direct consequence of the definition (\ref{stp}) 
together with \req{dg0}. The proof of the first property can be obtained by 
following the steps described in Appendix~E of 
Ref.~\onlinecite{catelani:aleiner:05}. 
We also notice that for the stationary state [see Appendix~\ref{app:stm}]:
\be\label{stmpr}
\lim_{t_{2}\to t_{1}} \St_{-} = 0.
\ee

The charge in the dot is given by:\footnote{To avoid rescaling the kinetic
equation, we switch back to units $\delta_1/\pi=1$.}
\be
q(t_1) = -\frac{e}{2} \lim_{t_{2} \to t_{1}} \Tr_s g(t_1,t_2), 
\ee
and taking the same limit of both sides of the kinetic equation (\ref{dkineq})
we obtain \req{chcon} with:\begin{subequations}\label{elcurrcc}
\bea
&& I_{\phantom{\alpha}} = I_L + I_R + I_-, \\
&& I_{\alpha}= e \sum_{n \in \alpha} \int\!\frac{d\epsilon}{2\pi} 
\left[ \frac{T_n}{2} \left( f_n (\epsilon) - g(\epsilon) \right) + 
I_{n}^{el} (\epsilon) \right], \quad\quad\quad \label{elcurrlr}\\
&& I_- = e \lim_{t_{2} \to t_{1}} \St_{-},
\eea
where $\alpha=L,R$ and
\be\label{iel}\begin{split}
I_{n}^{el} (\epsilon) = & - \frac{i}{8}\int\!\frac{d\omega}{2\pi} \,
T_n \left(1-T_n\right) K^R(\omega) \\ &
\times \left[ f_n(\epsilon-\omega) g(\epsilon) - 
g(\epsilon-\omega) f_n(\epsilon)\right].
\end{split}\ee
The contribution $I_-$ vanishes in the steady state, thanks to
\req{stmpr}, and will not be given any further consideration. We mention that the above expressions for the electric current can also be obtained by applying the gauge transformation \req{gts} to the current formula \req{apcur}.
\end{subequations}

We now turn to the energy conservation law, whose validity is based on the 
properties{\setlength\arraycolsep{0pt} \begin{subequations}
\bea\label{stinencon}
&& \lim_{t_{2}\to t_{1}} \frac{i}{2}\left( \partial_{t_{1}} - \partial_{t_{2}} \right)
\St^{in}\!\!= \partial_{t_{1}}  u_b (t_1) + \!\sumc I_n^{\varepsilon, in}(t_1), \quad\qquad\\
&& \lim_{t_{2}\to t_{1}} \left( \partial_{t_{1}} - \partial_{t_{2}} \right)
\St_+ = 0 \label{stppr2},
\eea\end{subequations}}
where 
\bes
I_n^{\varepsilon, in}(t_1) = T_n\!\left[ \frac{-i}{4\pi} \partial_t K^K 
\partial_t  - \frac{1}{8}\int\!dt_3 \left[ \partial_t K^R\right]_{1,3} 
g^{\phantom{n}}_{1,3}g^{\phantom{n}}_{3,1} \right] \\  
-\frac{1}{16}  T_n(1-T_n) \int\!dt_3
\left[ \partial_{t} K^R \right]_{1,3} 
\left[ g(f_n-g) +(f_n - g) g \right] ,
\end{split}
\ee
or after the Wigner transform
\bea\label{inelasticenergy}
&& I_n^{\varepsilon, in} = T_n \bigg[ \frac{i}{4\pi}\int\frac{d\omega}{2\pi} \omega^2 K^K (\omega) \\ && \qquad\qquad - \frac{1}{8}\int\frac{d\epsilon}{2\pi}\int\frac{d\omega}{2\pi} \omega \Im K^R(\omega) g(\epsilon - \omega)g(\epsilon)\bigg] 
\nonumber \\ 
&& -\frac{T_n(1-T_n)}{8}\! \int\frac{d\epsilon d\omega}{4\pi^2}\omega \Im K^R(\omega) g(\epsilon - \omega) (f_n(\epsilon) - g(\epsilon)).
\nonumber
\eea

As before, the last property can be proved straightforwardly, while
the derivation of the first one is delineated in Appendix~E of
Ref.~\onlinecite{catelani:aleiner:05}.
In order to write it in the above form we used the kinetic equations 
(\ref{boskineq}) and defined $u_b$ as
\be\label{ubdef}\begin{split}
u_b & = u_\rho - u_g, \\
u_\alpha & =  \frac{1}{2}{\cal L}^\alpha {\cal N}^\alpha \, , \quad \alpha=\rho,g.
\end{split}\ee
Below we will identify it as the contribution of collective excitations to 
the dot's energy. The separation in \req{stinencon} of a energy 
contribution from a energy current one may seem arbitrary; however there
are two independent tests of the validity of 
\reqs{inelasticenergy}-(\ref{ubdef}). First, since the leads' electrons are 
non-interacting, we can proceed similarly to the derivation of \req{apcur} for 
the electric current and find that the energy current is obtained by replacing
$e \to i(\partial_{t_{1}} - \partial_{t_{2}})/2$ in that equation; 
substituting in the resulting formula \reqs{s2} and (\ref{gts}), expanding to 
second order in $K_{\pm}$ 
and averaging over the fluctuating field, we again arrive at the results 
presented below, \reqs{encurrcc}. Second, the definition (\ref{ubdef}) of the 
collective excitations' energy is in agreement with the result of the 
thermodynamic calculation for such quantity, see Appendix~\ref{app:thermodyn}. 
Finally, we note that the property similar to \req{stmpr} holds in the steady 
state:
\be\label{stmpr2}
\lim_{t_{2}\to t_{1}} \left( \partial_{t_{1}} - \partial_{t_{2}} \right)
\St_{-} = 0.
\ee

The electrons' contribution to the dot's energy $u$ is
\be
u_e (t_1) = -\frac{i}{4} \lim_{t_{2} \to t_{1}}  
\left( \partial_{t_{1}} - \partial_{t_{2}} \right) \Tr_s g(t_1,t_2),
\ee
and from \req{dkineq} we arrive at
\be
\partial_t u_{tot} + I^\varepsilon = 0,
\ee
where the total energy $u_{tot}$ is the sum of the contributions of the 
electrons and collective excitations:
\be
u_{tot} = u_e + u_b.
\ee
The energy current $I^\varepsilon$ is:
\begin{subequations}\label{encurrcc}
\bea
&& I^\varepsilon = I^\varepsilon_L + I^\varepsilon_R + I^\varepsilon_- ,\\
&& I^\varepsilon_{\alpha}=  \sum_{n \in \alpha} \left[
\int\!\frac{d\epsilon}{2\pi} 
\, \epsilon \left[ \frac{T_n}{2} \left( f_n (\epsilon) - g(\epsilon) \right) + 
I_{n}^{el} (\epsilon) \right] + I_{n}^{\varepsilon, in} \right]\!,
\nonumber \\ && \\
&& I^\varepsilon_- = \frac{i}{2} \lim_{t_{2} \to t_{1}}\left( \partial_{t_{1}} - \partial_{t_{2}} \right) \St_{-} \, ,
\eea
where $\alpha=L,R$, $I_{n}^{el}$ is defined in \req{iel} and $I_{n}^{\varepsilon, in}$ in \req{inelasticenergy}.
Again, since $I^\varepsilon_-$ vanishes in the steady state -- see 
\req{stmpr2} -- we will not discuss it anymore.
\end{subequations}

\subsubsection{Final Form of the Kinetic Equation}

Closing this section, we present for completeness the final form of the
kinetic equation for the dot's distribution function $n(\epsilon)=\frac{1}{2} 
- \frac{1}{4} g(\epsilon)$. The collision integral is a functional of 
$n(\epsilon)$, the leads' distribution functions $\tilde{n}_n(\epsilon) = 
\frac{1}{2} - \frac{1}{4}f_n(\epsilon)$ and the bosonic distribution functions
$N^\alpha(\omega)$ defined in \req{bosdistfunc}:
\bea\label{gkineq}
&&\frac{\partial n(\epsilon)}{\partial t}  = 
\St \{ n,\tilde{n}_n, N^\alpha \},
\\ && \St = \St_{\tau} \{ n,\tilde{n}_n \} + \St^\rho\{n,\tilde{n}_n,N^\rho\}
 - \St^g\{n,\tilde{n}_n,N^g\}. \nonumber
\eea
In the second line, we separated in $\St$ three physically distinct 
contributions. The first term on the right hand side describes the relaxation 
of $n$ due to tunneling into and out of the contacts -- this mechanism is present
even for non-interacting electrons. The two other terms describe the 
interaction effects. Explicitly, they are given by
\be\label{nonint}
\St_{\tau} (\epsilon) = -\sum_n \frac{T_n}{2} \big[ n(\epsilon) - 
\tilde{n}_n(\epsilon) \big],
\ee
and
\bea\label{steb}
&& \St^\alpha(\epsilon) = -\delta_1 \left(1+F^\alpha \right)\sum_n
\int\! \frac{d\omega}{2\pi}\,\frac{1}{\omega} \\ 
&& \times \bigg\{ \frac{T_n}{4} \Big[ {\cal L}^\alpha(\omega) 
\tilde{\Upsilon}^\alpha_n (\epsilon ,\omega) + \tilde{\Upsilon}^\alpha_n(\epsilon,\omega) 
\bar{\cal L}^\alpha(\omega) \Big] \nonumber \\
&& +\frac{T_n(1-T_n)}{4} \Big[  \Re {\cal L}^\alpha(\omega)  \, \tilde{n}_n(\epsilon -\omega) \big( \tilde{n}_n(\epsilon) - n(\epsilon) \big) 
 \Big]\!\bigg\} ,\nonumber
\eea
where the propagators $\mathcal{L}^\alpha$, $\alpha = g,\rho$, are defined in \reqs{diffdef} and (\ref{Lc}), and we introduced the combination of distribution functions:
\be\begin{split}\label{combin}
\tilde{\Upsilon}^\alpha_n (\epsilon ,\omega) = & \big( N^\alpha(\omega) +1 \big)
\tilde{n}_n(\epsilon) \big( 1 - \tilde{n}_n(\epsilon -\omega )\big) \\
& - N^\alpha(\omega) \big( 1-\tilde{n}_n(\epsilon)\big)
\tilde{n}_n(\epsilon - \omega ) \, .
\end{split}\ee
 The bosonic distribution functions $N^\alpha(\omega)$, $\alpha = g,\rho$, were defined in \req{bosdistfunc} and they satisfy the kinetic equation
\bes\label{bosonkinet}
\frac{\partial N^\alpha(\omega)}{\partial t} = \frac{1+F^\alpha}{\omega} \sum_n \frac{T_n}{2} \int d\epsilon \bigg\{ \left[ \tilde{\Upsilon}^\alpha_n(\epsilon,\omega) + \Upsilon^\alpha(\epsilon,\omega)\right] \\
+ (1-T_n) \left[ n(\epsilon) - \tilde{n}_n(\epsilon) \right] \left[n(\epsilon-\omega) - \tilde{n}_n(\epsilon-\omega) \right] \bigg\},
\end{split}
\ee
which follows from \req{boskineq}. Here, $\Upsilon(\epsilon,\omega)$ is given by \req{combin} after the replacement $\tilde{n}_n \to n$. 
The combination (\ref{combin}) can be obtained by the standard argument for the creation
and annihilation of the one-particle excitations in the dot-lead system.
On the other hand the last line in \req{steb} may be understood as the renormalization of the scattering coefficients $T_n$ of the non-interacting collision integral \req{nonint}, due to interaction with the self-consistent potential in the dot.

\subsection{Triplet Channel}

The case of interaction in the triplet channel can be treated simlarly to the
singlet channel, the main difference being that the 
Hubbard-Stratonovich field, \req{hstruc}, is now a vector. Therefore the phase factors in the gauge transformation
must also possess this structure and in \req{gts} we substitute
\be
\hat{K} (t_i) \to \vec{\hat{K}}(t_i) \otimes \vec{\sigma}.
\ee
This transformation does not commute with the Zeeman energy term in
\req{tripusadel}; after the gauge transformation and expanding up to second 
order in $\vec{\hat{K}}$ this term becomes
\be
\frac{i}{2} E_Z^* \left[ \sigma^z ; \hat{g} \right] \to 
\frac{i}{2} E_Z^* \Big( \left[\sigma^z ;\hat{g} \right] + i \hat{K}^a 
\left[ \left[ \sigma^a;\sigma^z \right] ; \hat{g} \right] \Big),
\ee
where we sum over the repeated index $a=x,y,z$.
Note that the second order terms vanish identically.

The linear order equations for $K^a_\pm$ can be decoupled by using the
following basis for matrices in the spin space:
\bea
&& A = \frac{1}{2} \Tr_s (A) 1_s + A_m \sigma^m, \quad \left(A_m = \frac{1}{2} \Tr_s(\sigma^{-m} A) \right), \nl
&& \sigma^0 \equiv \sigma^z \, , \quad \sigma^{\pm 1} = \frac{1}{\sqrt{2}} \left(
\sigma^x \pm i\sigma^y \right).\label{news}
\eea
These matrices obey the following commutation relations:
\be
\left[ \sigma^0;\sigma^{\pm 1} \right] = \pm 2 \sigma^{\pm 1}\, , \quad
\left[ \sigma^{+1} ; \sigma^{-1} \right] = 2\sigma^0,
\ee
and all other commutators vanish. \footnote{The scalar product is now $\boldsymbol{a}\cdot
\boldsymbol{b}=a^zb^z+a^+b^-+a^-b^+$.} Then the solution for $K^m_-$, $m=0,\pm 1$ is
\be
K^m_- = - \bar{\cal L}^g_m h^m_-
\ee
(no summation over $m$), 
where ${\cal L}^g_m$ is obtained by shifting the frequency of ${\cal L}^g$ 
in \req{diffdef}:
\be
{\cal L}^g_m (\omega) = {\cal L}^g (\omega -mE^*_Z).
\ee
We can similarly obtain the expressions for $K^m_+$. In the new basis, the formula for the polarization operators $\Pi_{m,-n}$ is given by \req{polrt} with $(i,j)$ replaced by $(m,n)$, and we similarly obtain 
\be
\left[\hat{\Pi}^R_h\right]_{mn}(\omega) = -\frac{2}{\pi}\frac{\glr + i m E_Z^*}{-i(\omega - m E_Z^*) + \glr}\delta_{m,-n}, 
\ee
and using \req{tdes}
\be
\left[\D_h^R\right]_{mn} (\omega) = \frac{J_s}{2} \, \frac{- i \left(\omega - E_Z^*\right) + \glr}
{-i\omega + \left(1+\frac{J_s}{\delta_1}\right)\left(\glr + i m E_Z^*\right)}\delta_{m,-n},
\ee
for the interaction propagators. Using these expressions in the zero frequency
limit, together with \reqs{haverage}-(\ref{ezrenor}) we obtain the expression \req{EZ} for $E_Z^*$.  Finally for the retarded
propagator  we obtain $K^R_{mn} = K^R_m \delta_{m, -n}$, with
\be\label{krmn}
K^R_{m}(\omega) = \frac{\delta_1}{2}\frac{1}{-i\omega}\bigg[{\cal L}^g_m(\omega)
-\left(1+F^s \right) {\cal L}^s_m (\omega)  \bigg] ,
\ee
where
\be\label{lsm}
{\cal L}^s_m (\omega) = \frac{1}{-i\left(\omega - mE_Z\right) + 
\left(1 + F^s \right)\frac{1}{\tau}},
\ee
and 
\be
F^s \equiv \frac{J_s}{\delta_1}.
\ee
Similarly, $K^K_{m n}= K^K_m \delta_{m,-n}$ where 
\bea
K^K_m &=& - i\frac{\delta_1}{2} \frac{1}{\partial_t}
\bigg\{ \Big[{\cal L}_m^g {\cal N}_m^g +{\cal N}_m^g \bar{\cal L}_m^g\Big]
\\ && -
\left( 1+ F^s \right) \Big[{\cal L}_m^s {\cal N}_m^s
+ {\cal N}_m^s \bar{\cal L}_m^s \Big] \bigg\} \frac{1}{\partial_t}, \nonumber
\eea
and the density matrices ${\cal N}^\alpha_m$, $\alpha =g, s$, are defined similarly to \reqs{defnbos}. 
 
We could also repeat (with necessary modifications) the calculations of 
Sec.~\ref{sec:cc} and find the final form 
of the fermionic collision integral.  The kinetic equation \req{gkineq}, is modified by adding the following collision integrals to the right hand side: 
\be
\sum_{m=0,\pm 1} \left(\St^s_m - \St^g_m\right) ,
\ee 
where $\St^\alpha_m$ are given by \req{steb}  after replacing $\mathcal{L}^\alpha$ with $\mathcal{L}^\alpha_m$, and $N^\alpha$ with $N^\alpha_m$. The $N^\alpha_m$, $\alpha =g, s$,  are distribution functions for bosons with unit spin and are defined in terms of the density matrices ${\cal N}^\alpha_m$  as in \req{bosdistfunc}. They satisfy the kinetic equation \req{bosonkinet} after the replacement $N^\alpha \to N^\alpha_m$. 
It also follows that  the expressions for
the currents are obtained by replacing 
\be\label{substit}
K^R \to \sum_{m = 0,\pm 1} K^R_{m}
\ee
in \reqs{elcurrcc} and (\ref{encurrcc}).

\subsection{Cooper Channel}\label{sub3ofsec3}

For the interaction in the Cooper channel, we resort to a perturbative
approach which, in contrast to the charge and triplet channels, does not
start with a gauge transformation. Instead, we use 
the constraint (\ref{eq:cons}), to write the retarded and advanced Green 
functions as: 
\be\label{grcoop}
\boldsymbol{g}^R  =    \begin{pmatrix}
    \one -\frac{1}{2}\ff^R\!\cdot\bar{\ff}^R & \ff^R   \\
  \bar{\ff}^R    &  -\one+\frac{1}{2}\bar{\ff}^R\!\cdot\ff^R \end{pmatrix}_N,
\ee
\be
\boldsymbol{g}^A   =    \begin{pmatrix}
     -\one+\frac{1}{2}\ff^A\!\cdot\bar{\ff}^A & \ff^A   \\
  \bar{\ff}^A    & \one -\frac{1}{2}\bar{\ff}^A\!\cdot\ff^A\end{pmatrix}_N,
\ee    
and the Keldysh anomalous Green functions as:
\bea\label{fkcon}
\ff^K & = & -\frac{1}{2} ( \ff^R \!\cdot\!\bar{g}^K + g^K\!\cdot\!\ff^A),\nonumber\\
\bar{\ff}^K & = & + \frac{1}{2} (\bar{\ff}^R\!\cdot \!g^K + \bar{g}^K \!\cdot\!\bar{\ff}^A),
\eea
where  we use the notation \req{namstruc}.
In the normal phase the anomalous Green functions have only fluctuating parts proportional to $\Delta$, and hence the above expressions are valid up to second order in the fluctuating field. 
Using these formulas together with \req{s2}, and defining
\be\label{st12}
\Stb_1 \equiv  - i[ \hat{\boldsymbol{\Delta}}\, ;\hat{\boldsymbol{g}}], \qquad
\Stb_2 \equiv \sum_{n=1}^{\nc} \,[\,\hat{\boldsymbol{g}}\, ;\hat{\boldsymbol{g}}_n],
\ee
we can write the upper left GN component of the Keldysh part of the collision integral $\Stb_2$ as
\be \label{s22}
 (\Stb_2)_{11}^K = \sumc \left[\frac{T_n}{2}  (f_n^K - g^K) + \St^{el}_n +\St^{MT}_n\right] \ee
where we have identified the Maki-Thompson and elastic parts as
\bea
\St_n^{MT} &\equiv& \frac{1}{8}T_n^2 \ff^R(\bar{f}_n^K - \bar{g}^K)\bar{\ff}^A
\\
\St_n^{el\phantom{m}} &\equiv&  \frac{1}{8} T_n(T_n -1) \Big[ \ff^R\bar{\ff}^R 
(f_n^K - g^K) \nonumber \\
&& \quad + (f_n^K - g^K)\ff^A\bar{\ff}^A \Big].
\eea
We note that there is no term analogous to the Asla\-mazov-Larkin\cite{aslamazov:larkin} contribution -- 
this is due to the independence of the field $\Delta$ from the spatial
coordinate.
Similarly, the substitution of \reqs{grcoop}-(\ref{fkcon}) into $\St_1$ gives:
\bea \label{s1}
(\Stb_1)_{11}^K & = & - i\Delta_+\bar{\ff}^K -  i \ff^K\Delta_+^*-i \Delta_-\bar{\ff}^A - i \ff^R\Delta_-^* \nonumber\\
&=& - \frac{i}{2}\Delta_+\bar{\ff}^R g^K - \frac{i}{2}\Delta_+\bar{g}^K\bar{\ff}^A +\frac{i}{2}\ff^R\bar{g}^K \Delta_+^* \nl 
&& +\frac{i}{2}g^K\ff^A\Delta_+^* - i \Delta_-\bar{\ff}^A - i \ff^R\Delta_-^*
\eea

In order to average over the fluctuating field $\Delta$, we need to solve the equations
of motion for the $\ff$'s. Using \req{usacooper} for the $\ff^R$ component, we obtain
\bes\label{Fr}
\left(\partial_{t_{1}} - \partial_{t_{2}} + \frac{1}{\tau_*}\right)\ff^R + \frac{i}{2} E_Z^* \big\{\sigma^z;\ff^R\big\} = \\+2 i \Delta_+ \delta(t_1-t_2) +i g^K \Delta_-,
\end{split}\ee
where
\be\label{tsdef}
\frac{1}{\tau_*}= \glr + \frac{\rmg_h}{2},
\ee
and $\rmg_h$ was defined in \req{ghdef}. This modification of the escape rate is due to the last term on the left hand side of \req{usacooper}, which describes the breaking of time reversal invariance for the anomalous Green functions by the orbital magnetic field. It is clear from this equation that the $\ff$'s are diagonal in spin space, and if we decompose them according to 
\be
\ff \equiv \ff_+ \frac{1_s +\sigma^z}{2} + 
\ff_- \frac{ 1_s-\sigma^z}{2},
\ee
the solution to \req{Fr} can be expressed in terms of the cooperon $\C$ as:
\bea
\label{fff}
&&\ff^R_{\pm}(t_1-t_2) = i \,\C_{\pm}\!\left(\frac{t_1-t_2}{2}\right)\Delta_+\!\left(\frac{t_1+t_2}{2}\right) \\
&& + i\!\! \int\!dt_3 \, \C_{\pm}\!\left(\frac{t_1-t_2}{2} - t_3\right)\Delta_-\!\left(\frac{t_1+t_2}{2} - t_3\right) g^K\!\left(2 t_3
 \right), \nonumber
\eea
where
\be\label{cooperon}
\C_{\pm}(t) = \int\frac{d\epsilon}{2\pi}\frac{e^{-i\epsilon t}}{-i(\epsilon \mp E_Z^*) +\frac{1}{\tau_*}}.
\ee 
The $\bar{\ff}^R$ is obtained similarly and is given by \req{fff} after replacing $\Delta_\pm$ by $\mp\Delta_\pm^*$, and $g^K$ by $\bar{g}^K$. For the advanced components we can use \req{grelation} to write:
\bea
\ff^A_\pm (t_1-t_2) =  \ff^R_\mp (t_2-t_1); 
\eea
the same relation holds for the $\bar{\ff}$'s. From \req{grelation} we also 
have
\be
\bar{g}^K(t_1,t_2) = g^K(t_2,t_1), 
\ee
and similarly for $\bar{f}_n$ and $f_n$ (here and for the rest of this section
we drop the superscript $K$ from the dot and leads Keldysh Green functions).

Using these solutions for $\ff$'s and the propagators (\ref{cprop})
to average over the fluctuating fields, we arrive
at the following expressions for the singlet part of the Fourier transformed 
collision integral $\St_2$:
 \bea
&& \frac{1}{2} \mathrm{Tr}_s(\St_n^{MT}(\epsilon)) = \frac{1}{8} T_n^2 \!\int \frac{d\omega}{2\pi}\big( f_n(\omega -\epsilon) - g(\omega -\epsilon) \big) 
\nonumber \\ \label{koom} && \qquad
 \times\sum_{m=\pm 1} |\C_m|^2(2\epsilon-\omega)\J |\D_\Delta^R(\omega)|^2,
\eea
 \bea 
&& \frac{1}{2} \mathrm{Tr}_s(\St_n^{el}(\epsilon)) = \frac{1}{8}T_n(1-T_n) \!\int\! \frac{d\omega}{2\pi}\big(f_n(\epsilon)-g(\epsilon)\big) \nonumber \\&& \
 \times \bigg\{-2\sum_{m=\pm 1}\Re\, \C_m^2(2\epsilon -\omega)  \mathcal{J}(\omega - \epsilon,\omega) |\D_\Delta^R(\omega)|^2 \nonumber
\\ && \ +g(\omega-\epsilon)\sum_{m=\pm 1}\Im \left[ \C^2_m(\omega - 2\epsilon) 
\D_\Delta^R(\omega)\right]\bigg\},
\label{koom1} \eea
with the kernel $\J$ given by
\be\label{jkerdef}
\J = i \Pi_\Delta^K(\omega) + g(\epsilon) \Im\Pi_\Delta^R(\omega) \, .
\ee
To write the collision integral in this form we used the identity
\be\begin{split}
\big[ i \Pi_\Delta^K(\omega) &+ g(\epsilon) \Im\Pi_\Delta^R(\omega)\big] |D_\Delta^R(\omega)|^2 \\
&= i \D_\Delta^K(\omega) + g(\epsilon) \Im\D_\Delta^R(\omega),
\end{split}\ee
which follows from the Dyson equation (\ref{mdescoop}) for the propagators.
At $\omega$ much larger than $T$ or applied voltage $eV$, $i\D_\Delta^K(\omega) = -2 \mathrm{sgn}(\omega) \Im \D_\Delta^R(\omega)$, and the contribution from such large $\omega$'s in \req{koom} vanishes. This is a direct manifestation of the inelastic nature of Maki-Thompson processes.

By comparing
\reqs{apcur} and (\ref{s22}) with the definition of $\St_2$ [\req{st12}]
we find that the current in each channel is given by
\be\label{currncoop}
I_n = \frac{e}{2} \Tr_s \int\!\frac{d\epsilon}{2\pi} 
\left[\frac{T_n}{2} (f_n -g) +\St_n^{el} + \St_n^{MT} \right] 
\ee 
with the expressions (\ref{koom})-(\ref{koom1}) for the collision integrals.
We will use this result in Sec.~\ref{coopercond} to calculate the 
corresponding interaction correction to the conductance.

By repeating the above steps for the collision integral $\St_1$ we find:
\be\label{coopinel}
\frac{1}{2} \mathrm{Tr}_s(\St_1(\epsilon)) =  - \frac{1}{2}\sum_{m=\pm 1} \int \frac{d\omega}{2\pi} \Re(\C_m(2\epsilon-\omega)) \mathcal{K}(\epsilon,\omega) 
 \ee
where the kernel $\mathcal{K}(\epsilon,\omega)$ is 
\be\bs
\mathcal{K}(\epsilon,\omega)= &i \D_\Delta^K(\omega)\big[g(\epsilon)+g(\omega - \epsilon)\big]  \\ &+\Im \D_\Delta^R(\omega) \big[g(\epsilon)g(\omega-\epsilon) + 4\big].
\end{split}\ee

Having derived the collision integrals, we now find the explicit form of 
the propagators \req{cprop}, and polarization operators \req{polcoop}. 
Using Eqs.~(\ref{fkcon}) and (\ref{fff}) we obtain\begin{subequations}
\label{polracoop}
\bea
\Pi_\Delta^R (\omega)& = & 
\frac{-i}{2} \int \!\frac{d\epsilon}{2\pi} \sum_{m=\pm 1} \C_m(2\epsilon) 
g \Big(\frac{\omega}{2} -\epsilon\Big), \quad \\
\Pi_\Delta^A (\omega)& = & 
\frac{i}{2} \int \!\frac{d\epsilon}{2\pi} \sum_{m=\pm 1} \C_m(2\epsilon)
g \Big(\frac{\omega}{2} +\epsilon\Big),
\eea\end{subequations}
\be\label{polkcoop}
\Pi_\Delta^K(\omega) = -
\frac{i}{4}\!\int\!\frac{d\epsilon}{2\pi}\!\sum_{m=\pm 1}\!\C_m(2\epsilon) 
\Big[ g \Big(\frac{\omega}{2} -\epsilon\!\Big)
g \Big(\frac{\omega}{2}+\epsilon\!\Big) + 4\Big]\!.
\ee
In \req{polkcoop}, the last term in the brackets does not follow from the Usadel equation. It arises to compensate for the incorrect order of summation over orbital states $m$ and integration over $\epsilon$ in \req{polcoop}, similarly to the anomaly in charge channel [cf. the discussion after \req{polr}]. This ultraviolet factor can be obtained e.g. by explicit calculation in the original model. A more compact way, however, is to restore the correct term by requiring the fluctuation dissipation theorem: 
\be
\Pi_\Delta^K(\omega) = \coth{\left(\frac{\omega}{2T}\right)} \left[\Pi_\Delta^R (\omega) -\Pi_\Delta^A (\omega) \right],
\ee
to hold for the equilibrium ``distribution function'' $g(\epsilon) = 2 \tanh (\epsilon/2T)$. Furthermore the logarithmic divergence in \req{polracoop} is cut off by $|\epsilon| \lesssim E_T$ as it determines the validity of RMT. \cite{aleiner:brouwer:glazman:02}

In equilibrium, with $g(\epsilon)$ given by the right hand side of \req{fkeq} at temperature $T$ and $\mu=0$,  we can calculate $\Pi_\Delta^R$ explicitly by closing the integration contour in the upper half plane, obtaining a sum over the residues at the poles of $\tanh(\epsilon/2T)$. The result is  
\be\label{boob}
\Pi^R_{eq}(\omega) =\sum_{m=\pm 1} \Pi^R_{m}(\omega)= -
\frac{1}{2\pi}\sum_{m=\pm 1} \sum_{n=0}^{\tilde{N}} \frac{1}{n+\frac{1}{2} + z_m},
\ee
where 
\be\label{zmdef}
z_m \equiv \frac{-i(\omega - m E_Z^*) +\frac{1}{\tau_*}}{4\pi T}. 
\ee
The upper cutoff $\tilde{N}$ is approximately given by $E_T/2\pi T$ . We note that the coupling constant $J_c$ is also defined at the scale of $E_T$. 
We can express this sum in terms of the digamma function $\psi^{(0)}(x)$ 
\bea \label{kallepache}
-
2\pi\Pi_m^R(\omega) & =&  \psi^{(0)}\left(\frac{1}{2} + \frac{E_{T}}{2\pi T} + 
z_m\right) - \psi^{(0)}\left(\frac{1}{2}+z_m\right)\nonumber\\
& \simeq & \ln\left(\frac{E_{T}}{2\pi T}\right)- 
\psi^{(0)}\left(\frac{1}{2}+z_m\right) \, .
\eea
Defining the ``critical temperature'' 
\be\label{crittemp}
T_c \equiv \frac{2 \gamma E_{T}}{\pi}e^{\pi J_c^{-1}},
\ee
we can rewrite $\Pi^R_m$ as 
\be 
2\pi\Pi_{m}^R(\omega) = \pi J_c^{-1} + t +\psi^{(0)}\left(\frac{1}{2} + z_m\right)
-\psi^{(0)}\left(\frac{1}{2}\right),
\ee
where $t$ is the reduced temperature 
 \be
 t \equiv \ln{\frac{T}{T_c}}.
 \ee
Therefore, in equilibrium we find
 \be\label{propeq}
 \D^R_{eq} (\omega) = -\frac{\pi}{  t + \frac{1}{2}\sum_{m=\pm 1}\psi^{(0)}(\frac{1}{2} + z_m)-\psi^{(0)}(\frac{1}{2})}.
 \ee
Furthermore, in equilibrium we have the fluctuation-dissipation relation:
\be
\D^K_{eq}(\omega) = \coth{\left(\frac{\omega}{2T}\right)} 
\big[\D_{eq}^R(\omega)- \D_{eq}^A(\omega)\big] \, ,
\ee
as can be verified using \reqs{polracoop},
(\ref{polkcoop}) and (\ref{mdescoop}). We can define the bosonic distribution function of the fluctuating cooper pairs by generalizing this relation to the non-equlibrium case: 
\be\label{genfd}
\D^K_\Delta (\omega) = \left(2N^c(\omega) + 1 \right) \left[ \D^R_\Delta(\omega) - \D^A_\Delta (\omega)\right],
\ee
such that in equilibrium $N^c(\omega)$ is given by the Planck distribution \req{planck}.

The polarization operators must be recalculated in the non-equilibrium case,
as they depend on the distribution functions of the electrons. We do this for 
the DC case in Sec.~\ref{coopercond}. 

In closing this section we give the final form of the kinetic equation for interaction in the Cooper channel. Introducing $n(\epsilon)=\frac{1}{2} 
- \frac{1}{4} g(\epsilon)$ and $\tilde{n}_n(\epsilon) = 
\frac{1}{2} - \frac{1}{4}f_n(\epsilon)$ for distribution functions of the electrons in the dot and the leads respectively, and using \reqs{koom}--(\ref{coopinel}) and (\ref{genfd}) we arrive at
\bes
\frac{\partial n(\epsilon)}{\partial t}  =  -\sum_n \frac{T_n}{2} \big[ n(\epsilon) - 
\tilde{n}_n(\epsilon) \big] 
+ \St_{el} \{ n,\tilde{n}_n, N^c \} \\ + \St_{MT} \{ n,\tilde{n}_n, N^c \}+ \St_{1} \{ n,\tilde{n}_n, N^c \}.
\end{split}\ee
The collision integrals are given by
\be
\St_{el} = - \sum_n T_n (1-T_n) \big[ n(\epsilon) - 
\tilde{n}_n(\epsilon) \big] \int\frac{d\omega}{2\pi} \mathcal{M}(\epsilon, \omega),
\ee
where
\bes
\mathcal{M}(\epsilon, \omega) = \left[ N^c(\omega) + n(\epsilon)\right] \sum_{m=\pm1}\Re \, \C_m^2 (2\epsilon - \omega) \Im \D^R_\Delta(\omega) \\ - \frac{1}{2} n(\omega -  \epsilon) \sum_{m=\pm 1}\Im \left[ \C^2_m(\omega - 2\epsilon) \D^R_\Delta(\omega)\right],
\end{split}\ee
and 
\bea
\St_{MT} &=& \sum_n \frac{T_n^2}{2} \int\frac{d\omega}{2\pi} \sum_{m=\pm 1} |\C|_m^2(2\epsilon - \omega) \Im \D^R_\Delta(\omega) \nonumber \\ 
&& \times \left[ \tilde{\Psi}_n(\epsilon ,\omega) -  \Psi(\epsilon ,\omega) \right],
\eea
\bes
\St_1 = - 2 \int\frac{d\omega}{2\pi} \sum_{m=\pm 1} \Re \, \C_m (2\epsilon - \omega) \Im \D^R_\Delta(\omega) \Psi (\epsilon ,\omega).
\end{split}\ee
Here we defined
\be\begin{split}\label{kharmar}
\Psi (\epsilon ,\omega) = & - \big( N^c(\omega) +1 \big)
n(\epsilon) n( \omega - \epsilon ) \\
& + N^c(\omega) \big( 1-n(\epsilon)\big)
\big(1- n( \omega - \epsilon )\big) \, ,
\end{split}\ee
and $\tilde{\Psi}_n$ is obtained by replacing $n(\omega - \epsilon)$ with $\tilde{n}_n(\omega - \epsilon)$ in \req{kharmar}. The bosonic distribution function $N_c(\omega)$ is in turn given in terms of the fermionic distribution function $n(\epsilon)$ through
\be
2 N_c(\omega) + 1 = \frac{\Pi_\Delta^K(\omega)}{ \Pi_\Delta^R(\omega)-\Pi_\Delta^A(\omega)},
\ee
and the expressions (\ref{polracoop})--(\ref{polkcoop}) for the polarization operators.

\section{Calculations of the transport coefficients}\label{sec:calc}

In this section we explicitly calculate the interaction corrections
to the electrical and thermal conductances by solving the kinetic equations.
We first consider the singlet and triplet channel corrections to the 
electrical conductance and then to the thermal conductance. Finally, 
we calculate the Cooper channel correction to the electrical conductance.

\subsection{Electrical conductance: singlet and triplet channel}

In the presence of a bias voltage, the energy levels in the leads are shifted 
by the voltage times the electron's charge, so that the 
nonequilibrium ``distribution functions'' in the leads, \req{fkeq}, 
become 
\bea\label{leads}
f_{R(L)} (\epsilon) =  2 \tanh\left(\frac{\epsilon - e V_{R(L)}}{2 T}\right), 
\eea
with $V_{R(L)} = \pm V/2$.  For the future use we introduce
\be\label{delf}
\Delta f (\epsilon) \equiv f_L(\epsilon) - f_R(\epsilon).
\ee

In the steady state 
we have $I_L = - I_R \equiv I$, so we can write
\be
I = x I_L + (x-1) I_R,
\ee
for any constant $x$. In terms of the current in each channel
\bes\label{current2}
I  =\sumc \Lambda_n I_n 
\end{split}\ee
where we choose the constants to be 
\begin{equation}
\Lambda_n \equiv \left\{ \begin{array}{ll} \frac{\rmg_R}{\rmg_L+\rmg_R} & 1\leq n \leq N_L, \\ & \\  - \frac{\rmg_L}{\rmg_L+\rmg_R} & N_L + 1\leq n \leq \nc,
\end{array}\right. 
\end{equation}
with $\rmg_L, \rmg_R$ defined in \req{glgr}.
This choice will simplify calculations as we make use of the identity 
$\sum_n\Lambda_n T_n= 0 $.  

Let us calculate the singlet channel correction 
first. Using \req{elcurrlr} we have
 \be
 I = e \sumc \Lambda_n \int\!\frac{d\epsilon}{2\pi} 
\left[ \frac{T_n}{2} \big( f_n (\epsilon) - g(\epsilon) \big) + 
I_{n}^{el} (\epsilon) \right], 
\ee
with $I_{n}^{el}$ given in \req{iel}.
By shifting the integration variable $\epsilon$ by $\omega$ in the first 
term in square brackets in \req{iel},  
we can rewrite $I$ as:
\bes  \label{current3}
I =& e  \int\!\frac{d\epsilon}{2\pi} \sum_n \Lambda_n  \bigg\{ \frac{T_n}{2}\big(f_n(\epsilon) - g(\epsilon)\big)\\ &+\frac{1}{4}T_n(1-T_n) f_n(\epsilon) K(\omega) g(\epsilon-\omega) \bigg\},
\end{split}\ee
where
\be
K(\omega) \equiv -\Im K^R(\omega) = \Im \big[\! 
\left(\mathcal{L}^g\right)^2(\omega) \D^R(\omega) \big],
\ee
and we used \req{krldl} in the last identity; in this form we can recognize
that our expression for the current has the structure similar to the 
one found in Ref.~\onlinecite{altshuler:aronov:book} for higher dimensional 
systems. However, for practical purposes, we will use the following identity:
\be\label{chin}
K(\omega) = \frac{\pi\tau}{2}\Im \big[\mathcal{L}^g(\omega) - \mathcal{L}^\rho(\omega)\big],
\ee
where $\mathcal{L}^g$ and $\mathcal{L}^\rho$ were defined in \reqs{diffdef} and (\ref{Lc}) respectively.

 We note that since $\sum_n \Lambda_n T_n g = 0$, the interaction correction to $g$ does not enter the expression for the current to first order in the interaction propagator. 
So for the purpose of calculating the DC current, we can calculate $g$ 
to zeroth order by equating to zero the first term on the right hand side
of \req{dkineq} and we find:
\be\label{g}
g (\epsilon)= \frac{\mathrm{g}_L f_L(\epsilon) + \rmg_R f_R(\epsilon)}{\rmg_L + \rmg_R}.
\ee
with $\rmg_L, \rmg_R$ defined in \req{glgr}.

We write the differential conductance as
\be\label{dimless}
G = \frac{d I}{d V} = G_0 + \Delta G,
\ee
where $G_0$ and $\Delta G$ are the classical conductance and the interaction correction, originating from the first and second term in curly brackets
in \req{current3} respectively.
Substituting \req{g} into \req{current3} and using the definition 
(\ref{dimless}) we obtain
\be
G_0 =  \frac{e^2}{2} \frac{\rmg_L \rmg_R}{\rmg_L + \rmg_R}   \frac{\partial}{\partial \tilde{V}} \!\int \frac{d\epsilon}{2\pi}\Delta f (\epsilon)= \frac{e^2}{\pi} \frac{\rmg_L \rmg_R}{\rmg_L + \rmg_R}, 
\ee
and
\bes \label{current4}
\Delta G= &\frac{\rmg_R^2 \rmh_L +  \rmg_L^2\rmh_R}{(\rmg_L + \rmg_R)^2} \frac{e^2}{4}   \\ & \times  \frac{\partial}{\partial \tilde{V}}\!\int\! \frac{d\epsilon}{2 \pi} \!\int\! \frac{d\omega}{2 \pi} \Delta f(\epsilon) K(\omega) g(\epsilon - \omega),
\end{split}\ee 
where the form factors $\rmh_L$ and $\rmh_R$ are defined in \req{hlr}, and \be\nonumber
\tilde{V} \equiv eV.
\ee

We define the shorthand notation
\be
a \K b \equiv \int \frac{d\epsilon}{2 \pi} \int \frac{d\omega}{2 \pi} a(\epsilon) K(\omega) b(\epsilon - \omega),
\ee
and notice that since $K$ is an odd function of $\omega$, we can further 
simplify \req{current4} into 
\be\label{deltags}
\Delta G=  \frac{e^2}{\pi} \frac{\rmh_L \rmg_R^2 + \rmh_R \rmg_L^2}{(\rmg_L + \rmg_R)^2} \frac{\pi}{4} f'_R\K f_L
\ee
where here and below a prime denotes derivation with respect to the energy 
variable.
The integration over $\epsilon$ can be readily performed
and after a partial integration over $\omega$ we arrive at
\be\label{shitake}
\frac{\pi}{4} f'_R\K f_L = 2 T \int \frac{d\omega}{2\pi} K'(\omega)\frac{  \omega-\tilde{V}}{2 T}\coth{\frac{ \omega-\tilde{V}}{2 T}}.
\ee
Using the identity (\ref{chin}) together with the definitions Eqs.~(\ref{diffdef}) and (\ref{Lc}), we integrate over $\omega$ by closing the contour in the upper half plane, obtaining a sum over the residues of the poles of $\coth{(\omega/2T)}$, which is expressible in terms of polygamma functions. The result is:
\bes
&\frac{\pi}{4} f'_R\K f_L =  \\ &\frac{\tau}{2} \Re \!\left[\Psi\!\left( \frac{1 - i \tilde{V}\tau}{2\pi\tau T}\right)-\Psi\!\left(\frac{1+\frac{4 E_c}{\delta_1} - i \tilde{V}\tau}{2\pi \tau T}\right)\right],
\end{split}\ee
where
\be
\Psi(z)\equiv z\psi^{(1)}(z)+\psi^{(0)}(z).
\ee
and $\psi^{(i)}(z)$ is the $i-$th derivative of the digamma function.
Substitution into \req{deltags} yields \req{condu} for the singlet channel 
correction to the electrical conductance (where we have also restored 
dimensionful units).

As for the triplet channel, the result is found by using the 
replacement in \req{substit} together with the definitions (\ref{krmn}) and 
(\ref{lsm}). Then \req{chin} becomes:
\be
K(\omega) \to \sum_{m=0,\pm 1} \frac{\pi}{2} \frac{1}{1/\tau + imE_Z^*}
\Im \left[\mathcal{L}^g_m(\omega) - \mathcal{L}^s_m(\omega)\right]
\ee
and repeating the above steps 
we obtain \req{tripl} for the triplet channel correction $\Delta G_s$. 

\subsection{Thermal conductance}

The thermal conductance is found by assuming that there is a 
temperature difference $\delta T$ between the right and left leads, so that 
the distribution functions, cf. \req{fkeq}, are given by 
\be
f_{R(L)}(\epsilon)=  2 \tanh\left(\frac{\epsilon}{2 T \pm \delta T}\right),
\ee
where $+$ ($-$) signs corresponds to right (left) lead.
At lowest order, $g$ is still given by the relation (\ref{g}). 

We write the linear thermal conductance as 
\be
\kappa \equiv \frac{I^{\varepsilon}}{\delta T}=\kappa_0 + \Delta\kappa_{el} + 
\Delta\kappa_{in},
\ee 
where $\kappa_0$ is the classical thermal conductance and $\Delta\kappa_{el}$ 
and $\Delta\kappa_{in}$ are the contributions of the elastic and inelastic 
processes described respectivley by the collision integrals in 
\reqs{stelch} and (\ref{stinch}). The calculation is similar to the one 
in the previous section: we can write the steady state
energy current as $\sum_n \Lambda_n I_n^{\varepsilon}$ and from \reqs{inelasticenergy} and (\ref{encurrcc}) we obtain 
\begin{subequations}
\bea
I^\varepsilon &=& I^\varepsilon_0 + I^\varepsilon_{el} + I^\varepsilon_{in} \\ 
I^{\varepsilon}_0&=&\frac{1}{2}\sumc \Lambda_n T_n 
\int\!\frac{d\epsilon}{2\pi} \, 
\epsilon \big( f_n (\epsilon) - g(\epsilon) \big), \\ 
I^{\varepsilon}_{el} &= & - \frac{i}{8}\sumc \Lambda_n T_n 
\left(1-T_n\right)\int\!\frac{d\epsilon}{2\pi} \int\!\frac{d\omega}{2\pi} \,
 K^R(\omega) \nonumber \\ &&
\times \epsilon \big[ f_n(\epsilon-\omega) g(\epsilon) - 
g(\epsilon-\omega) f_n(\epsilon)\big], \\
I^{\varepsilon}_{in} &=& - \frac{1}{16}\sumc \Lambda_n T_n(1-T_n) \int\!\frac{d\epsilon}{2\pi}  \int\!\frac{d\omega}{2\pi i} \,
 \omega K^R\!(\omega) \nonumber \\
&& \times \big[ g(\epsilon -\omega)
\big(f_n-g\big)(\epsilon) + \big(f_n-g\big)(\epsilon -\omega) g(\epsilon) \big]. \nl
\eea\end{subequations}
We note that the part of $I^{\varepsilon}_{in}$ proportional to $T_n$'s [see \req{inelasticenergy}], vanishes in the DC case after summing over the channels due to the identity $\sum_n \Lambda_n T_n = 0$, and was therefore omitted here.
Summing over the channels explicitely
and expanding the distribution functions to linear order in $\delta T$, we 
find
\be\label{kzero}
\kappa_0= \frac{\pi G_0}{2e^2}\int\!\frac{d\epsilon}{2\pi} \frac{\epsilon^2}{2T} \frac{\partial f_0}{\partial \epsilon}
=\frac{\pi^2}{3 e^2}T G_0 \, ,
\ee
in agreement with the Wiedemann-Franz law, and\begin{subequations}\label{delk}
\bea
\Delta \kappa_{el}&= & \frac{A}{4\tau}\int\!\frac{d\omega}{2\pi i} \,\int\!\frac{d\epsilon}{2\pi} \, \mathcal{I}_{-}(\omega,\epsilon)\, \epsilon \, K^R(\omega),\\
\Delta \kappa_{in}&=&-\frac{A}{8\tau} \int\!\frac{d\omega}{2\pi i}\int\!\frac{d\epsilon}{2\pi} \, \mathcal{I}_{+}(\omega,\epsilon)\, \omega\, K^R(\omega), 
\quad\quad
\eea\end{subequations}
where 
\be\label{A}
A= \frac{\rmh_L \rmg_R^2 + \rmh_R \rmg_L^2}{(\rmg_L + \rmg_R)^3} ,
\ee
is the same form factor that appears in equation (\ref{current4}),
and $\mathcal{I}_{\pm}$ are the following combinations
of distribution functions:
\be
\mathcal{I}_{\pm}(\omega,\epsilon) = \frac{1}{\delta T} \left[f_0(\epsilon)
\Delta f(\epsilon - \omega) \pm  f_0(\epsilon - \omega)
\Delta f(\epsilon)\right] \, .
\ee
Note that at linear order
\be\label{linearf}
\Delta f(\epsilon) \simeq \frac{\delta T}{2T} \epsilon 
\frac{\partial f_0}{\partial \epsilon},
\ee
where $f_0$ is the equilibrium Green function given in \req{fkeq} with temperature $T$ and $\mu=0$.

In \reqs{delk}, we perform the integration over $\epsilon$:
\begin{subequations}
\bea
\label{det}\int\!\frac{d\epsilon}{2\pi} \, \epsilon \, \mathcal{I}_{-}(\omega,\epsilon) &=& \frac{4}{2\pi  T}\frac{\omega^3}{6}\frac{\partial}{\partial \omega}\coth{\frac{\omega}{2T}} \\
&& +\frac{8T}{\pi}\frac{\pi^2}{3}\frac{\partial}{\partial \omega}\left(\omega\coth{\frac{\omega}{2T}}\right), \nl
\int\!\frac{d\epsilon}{2\pi} \, \mathcal{I}_{+}(\omega,\epsilon)& = & - \frac{4}{2\pi  T}\omega^2\frac{\partial}{\partial \omega}\coth{\frac{\omega}{2T}}.
\eea\end{subequations}
After substitution in \req{delk}, the last term in the right hand side of \req{det} gives a term proportional (after intergrating $\omega$ by parts) to the integral \req{shitake} appearing in the calculation of the electrical conductance, and so
contributes to $\Delta \kappa_{el}$ a term that obeys 
the Wiedemann-Franz law; the first term, on the other hand, gives a 
contribution proportional to the inelastic correction:\begin{subequations}
\label{deltakappa}\bea
\Delta \kappa_{el} & = & \frac{\pi^2}{3e^2}T\Delta G_c - \frac{A}{3} J_{in}
\, , \\
\Delta \kappa_{in} & = & -A J_{in},
\eea\end{subequations}
where 
\bes
J_{in} &\equiv - \frac{1}{4\pi\tau T}\int\!\frac{d\omega}{2\pi i}  \omega^3\frac{\partial}{\partial \omega}\coth{\frac{\omega}{2T}} \, K^R(\omega).
\end{split}\ee
In this expression, the final integration over $\omega$ can be done similar to \req{shitake} by closing the contour in the upper half plane,
and the result can be expressed in terms of polygamma functions:
\be
J_{in} = \frac{\pi T}{12} \left[g_1\left(2\pi\tau T\right) 
- g_1\left(\frac{2\pi\tau T}{1+\frac{4E_c}{\delta_1}}\right) \right],
\ee
where\cite{catelani:aleiner:05}
\be
g_1(x) = \frac{6}{x^3}\psi^{(1)}\left(\frac{1}{x}\right)-\frac{6}{x^2}-\frac{3}{x}.
\ee

Equation (\ref{kzero}) together with \reqs{deltakappa} gives the 
singlet channel part of the result for the thermal conductance reported in
Sec.~\ref{sec:thermcondres}. Evaluation of the triplet channel contribution
is straightforward (as explained at the end of the previous subsection)
and leads to \req{thermtripl}.

\subsection{Electrical conductance: Cooper channel}\label{coopercond}

The evaluation of the Cooper channel correction to the electrical conductance 
is based on the current formula (\ref{currncoop}).
Writing the DC current as in \req{current2} and summing over $n$, the 
contribution of the first term in square brackets in \req{currncoop} gives the 
classical
conductance $G_0$,  and the contribution of the other terms gives the Cooper channel correction which can be written as:
\be
\Delta G_{Cooper} = \Delta G_{el} + \Delta G_{MT},
\ee
where
\bea\label{khaar}
\Delta G_{el} \ \; \, &=& \frac{e^2}{\pi} A \left[ J_{DoS_{1}} + J_{DoS_{2}} \right], \\
\Delta G_{MT} &=& \frac{e^2}{\pi} B J_{MT}, \label{dgmtj}
\eea
with the form factor $A$ defined in \reqs{A} and
\be\label{B}
B=\frac{\rmg_L\rmg_R}{(\rmg_L + \rmg_R)^2} - A,
\ee 
which are characteristic of elastic and inelastic contributions respectively,
and we introduced
\be\label{IMT1}\bs
J_{MT}  \equiv &
\frac{2}{\tau}\frac{\pi}{8}  \frac{\partial}{\partial \tilde{V}} \!\int\!\frac{d\omega}{2\pi} \! \int\!\frac{d\epsilon}{2\pi}  \sum_{m=\pm 1} |\C_m|^2(2\epsilon) \\ 
& \times \Delta f \left(\frac{\omega}{2} + \epsilon \right) 
\mathcal{J}\left(\frac{\omega}{2}-\epsilon,\omega\right) |\D_\Delta^R|^2(\omega),
\end{split}\ee
\be\label{IRe1}\bs
J_{DoS_{1}} \equiv &
- \frac{2}{\tau}\frac{\pi}{8}  \frac{\partial}{\partial \tilde{V}} \!\int\!\frac{d\omega}{2\pi} \! \int\!\frac{d\epsilon}{2\pi}  \sum_{m=\pm 1}  2 \Re\,\C_m^2(2\epsilon) \\ 
& \times \Delta f \left(\frac{\omega}{2} + \epsilon\right) 
\mathcal{J}\left(\frac{\omega}{2}-\epsilon,\omega\right) |\D_\Delta^R|^2(\omega),
\end{split}\ee
\bes\label{dooos}
 J_{DoS_{2}} \equiv & \frac{2}{\tau}\frac{\pi}{8}  \frac{\partial}{\partial \tilde{V}} \int\!\frac{d\epsilon}{2\pi} \! \int\!\frac{d\omega}{2\pi} 
\Delta f \left(\frac{\omega}{2}-\epsilon \right) \\
& \times g \left(\frac{\omega}{2} + \epsilon\right)  
\Im \bigg[ \sum_{m=\pm 1} \C^2_m(2\epsilon) \D_\Delta^R(\omega)\bigg].
\end{split}\ee
The kernel $\J$ and the function $\Delta f(\epsilon)$ are defined in 
\reqs{jkerdef} and (\ref{delf}) respectively.

To proceed further, we need the explicit form of the polarization operators 
in the presence of the bias voltage, as they determine the the kernel $\J$. 
In this case to zeroth order in interaction we have [see \req{g}]: 
\be\label{gg2}
\frac{1}{2} g^K(\epsilon)= \sum_{\alpha=\pm 1} \rmg_\alpha \tanh
\left(\frac{\epsilon - \alpha \tilde{V}/2}{2T}\right),
\ee
where $\rmg_{\pm1}$ are defined as
\be\label{gpm}
\rmg_{+1} \equiv \frac{\rmg_R}{\rmg_L + \rmg_R} \, , \quad
\rmg_{-1} \equiv \frac{\rmg_L}{\rmg_L+\rmg_R}.
\ee
Using the formula (obtained similarly to \req{kallepache})
\be\label{ep1}
\int\! d\epsilon \,\Re\,\C_m(2\epsilon) \tanh\!\left(
\frac{\epsilon - \omega/2}{2T}\right)  = 
\Im\,\psi^{(0)}\!\left(\frac{1}{2}+z_m\right),
\ee
with $z_m$ given in \req{zmdef}, and the definition
\be\label{khay}
\ipsi{i}{m}{\alpha} \equiv  \frac{1}{\pi} \Im\,\psi^{(i)}\!\left(\frac{1}{2}+
\frac{\frac{1}{\tau_*} + i (\omega-m E_Z^* -\alpha \tilde{V})}{4\pi T}\right),
\ee
with $\psi^{(i)}$ the $i$-th polygamma function, 
from \req{polkcoop} we obtain
\be\label{pkcf}
i\Pi_\Delta^K\!(\omega) = 
\sum_{m,\alpha,\beta} \rmg_\alpha \rmg_\beta \coth{\left(\frac{2\omega - (\alpha + \beta) \tilde{V}}{4T}\right)}\ipsi{0}{m}{\alpha} \, .
\ee
Here and below, the indices $m$, $\alpha$ and $\beta$ are summed over 
$\pm 1$; the quantities $\tau_*$ and $E_Z^*$ are defined in \reqs{tsdef} 
and (\ref{EZ}).
Using again \reqs{ep1}-(\ref{khay}), \reqs{polracoop} give: 
\be\label{impr}
\Im \, \Pi_\Delta^R(\omega)  =  -\frac{1}{2} \sum_{m,\alpha} 
\rmg_\alpha \ipsi{0}{m}{\alpha},
\ee
Substituting the above results into \req{jkerdef} we find:
\bea\label{brak}
\J & =& 
\sum_{m,\alpha,\beta} \rmg_\alpha \rmg_\beta \ipsi{0}{m}{\alpha} \\ &
\times & \left[\coth\!\left(\frac{2\omega - (\alpha + \beta) \tilde{V}}{4T}\right) - 
\tanh\!\left(\frac{2\epsilon - \beta \tilde{V} }{4T}\right)\!\right]\! . 
\nonumber
\eea

What remains to be done, are the two integrals over $\omega$ and $\epsilon$
in \reqs{IMT1}-(\ref{dooos}); the latter can be evaluated exactly, while the 
former only approximately. Due to their similarity, we consider 
$J_{MT}$ and $J_{DoS_{1}}$ together in the next subsection, deferring the 
calculation of $J_{DoS_{2}}$ to a later subsection.

\subsubsection{Maki-Thompson and $DoS_{1}$ parts}

Substituting \req{brak} into \reqs{IMT1} and
using \req{ep1} together with the identity:
\be
\frac{1}{\tau_*} |\C_m|^2(2\epsilon) = \Re\, \C_m(2\epsilon),
\ee
the result of the integration over $\epsilon$ is:
\bea\label{hamaan}
&& J_{MT} = \sum_{m,n}\sum_{\alpha,\beta} \alpha \rmg_\alpha \rmg_\beta \frac{\pi}{4}  \frac{\partial}{\partial \tilde{V}} \int\!\frac{d\omega}{2\pi}  |\D_\Delta^R|^2(\omega)
\label{jmt11} \\ 
&& \times\left[\coh{2\omega - (\alpha + \beta) \tilde{V}} - 
\coh{2\omega - (  \beta-\alpha) \tilde{V}}\!\right] \nonumber \\ && 
\times \ipsi{0}{m}{\alpha} \left(\ipsi{0}{n}{-\alpha} + \ipsi{0}{n}{\beta}\right), \nonumber
\eea
where $\rmg_\alpha$ and $\ipsi{i}{m}{\alpha}$ were defined in \reqs{gpm} and (\ref{khay})  respectively, and the subscripts $m$, $n$, $\alpha$ and $\beta$ are summed over $\pm 1$.  
Moreover since 
\be\label{sonbol}
-\Re\, \C_m^2(2\epsilon) = \frac{\partial}{\partial (\tau_*^{-1})} \Re\,\C_m(2\epsilon),
\ee
the expression for $J_{DoS_{1}}$ is found by replacing the $\chi^n_0$'s with 
$\chi^n_1/(2\pi\tau_* T)$ in \req{hamaan}.
For convenience,  in both $J_{MT}$ and $J_{DoS_{1}}$, we separate two contributions
with different dependences on $\rmg_{L,R}$ (and we drop a contribution
odd in $\omega$ which vanishes upon integration):
\be\label{ja}\bs
J_{a} = \frac{2 \rmg_L \rmg_R}{(\rmg_L+\rmg_R)^2} J^1_{a} + 
\frac{\rmg_L^2 + \rmg_R^2}{(\rmg_L+\rmg_R)^2} J^2_{a} 
\end{split}\ee
where $a=MT,DoS_{1}$. For $a=MT$ we have
\bea
\label{Jmt1}
J^1_{MT} &=& \frac{\pi}{4}  \frac{\partial}{\partial \tilde{V}} \int\!\frac{d\omega}{2\pi}  |\D_\Delta^R|^2(\omega)\sum_{m,n,\alpha}\ipsi{0}{m}{\alpha} \ipsi{0}{n}{-\alpha}
\nonumber \\
&& \times \alpha\left[\coth\left(\frac{\omega  -\alpha \tilde{V}}{2T}\right) - 
\coth\left(\frac{\omega}{2T}\right)\right]
\\
\label{Jmt2}
J^2_{MT} &=&   \frac{\pi}{4}  \frac{\partial}{\partial \tilde{V}} \!\int\!\frac{d\omega}{2\pi}  |\D_\Delta^R|^2(\omega)\sum_{m,\alpha}\ipsi{0}{m}{\alpha}\frac{1}{2}\sum_{n,\beta} \ipsi{0}{n}{\beta}\hspace{0cm} \nonumber \\ &&
\times \alpha\left[\coth\left(\frac{\omega  -\alpha \tilde{V}}{2T}\right) - 
\coth\left(\frac{\omega}{2T}\right)\right],
\eea
and, as explained above, the formulas for $a=DoS_{1}$ are found by replacing $\chi_0^n$ by $\chi_1^n/(2 \pi \tau_* T)$.

Up to now, no approximation has been made. However 
the validity of the RMT for the metallic dots requires all the energy 
scales [$T$,$\tilde{V}$,$1/\tau_*$,$E_Z^*$] to be much smaller than $T_c$; we can then neglect the energy dependence of the 
propagator and approximate 
\be\label{propapprox}
\D_\Delta^R (\omega) \simeq \D_\Delta^R(0) \simeq \frac{\pi}{\varepsilon}
\ee
with $\varepsilon$ defined in \req{redtemp}. The last approximation is 
valid with logarithmic accuracy and can be verified by noticing
that the dependence of the polarization operator on the relevant quantities is 
the same as in \req{impr} (dropping ``$\Im$'' on both sides of the equation) 
and using \req{propeq}.
Even with this approximation, the integrals can not be calculated in closed 
form and we must resort to further approximations valid in different limits. 
In particular, we will consider the low and high temperature regimes -- the 
transition between the two occurs at $T\sim 1/\tau_*$, with
$\tau_*$ defined in \req{tsdef}.

In the low temperature limit $T \ll 1/\tau_*$, we can use in
\req{khay} the following asymptotics for the polygamma functions:
\bea\label{arc}
\Im \, \psi^{(0)}\left(\frac{1}{2} +\frac{\frac{1}{\tau_*} + i \omega}{4\pi T}\right) &\approx & \arctan{\tau_* \omega}\\
\Im \, \psi^{(1)}\left(\frac{1}{2} +\frac{\frac{1}{\tau_*} + i \omega}{4\pi T}\right)&\approx & -4 \pi T\tau_* \frac{\tau_*\omega}{1+(\tau_*\omega)^2}. \quad\quad
\eea
We further distinguish two cases: ``low'' and ``high'' voltage, when
$\tilde{V}\ll 1/\tau_*$ and $\tilde{V}\gg T$ respectively; it is evident that the two
conditions are not mutually exclusive and the results derived below
must agree with each other at intermediate voltages. 

If the voltage is small, $\tilde{V} \ll 1/\tau_*$, we can further expand 
$\ipsi{i}{n}{\alpha}$
in $\tau_* \tilde{V}$ and $\tau_*\omega$ as well (as $|\omega|$ is itself limited 
by $|\tilde{V}|$) to obtain
\bea\label{c0lta}
\sum_n \ipsi{0}{n}{\alpha} &\approx& \frac{2}{\pi} \frac{\tau_*(\omega -\alpha \tilde{V})}{1 + (\tau_* E_Z^*)^2}.
\eea
Moreover in this approximation we have
\be
\sum_n \ipsi{1}{n}{\alpha} \approx -4\pi T\tau_* \frac{1- (\tau_*E_Z^*)^2}
{1+ (\tau_*E_Z^*)^2} \sum_n \ipsi{0}{n}{\alpha}, 
\ee
so that in this regime
\be
J_{DoS_{1}} = -2 \frac{1- (\tau_*E_Z^*)^2}{1+ (\tau_*E_Z^*)^2} J_{MT},
\ee
and we only need to calculate $J_{MT}$.

Inserting \req{c0lta} into \req{Jmt1} and performing the integral over 
$\omega$ we find
\bes\label{j1mt1}
J^1_{MT}  
&\approx  \left[ \frac{2}{\varepsilon(1 + (\tau_* E_Z^*)^2)}\right]^2 
\left[ (\tau_* \tilde{V})^2 + \frac{\pi^2}{3} (\tau_* T)^2\right],
\end{split}\ee
where we used the identity
\bea 
&&\frac{\partial}{\partial y} \int\! dx \left[
\coth (x  - y) - \coth(x+y)\right] 
(x^2 - y^2) \nonumber
\\ && \quad\quad = 8 y^2 + \frac{2\pi^2}{3}.
\eea

By repeating the above steps for \req{Jmt2} we find
\be\label{j2mt1}
J^2_{MT} \approx   \left[ \frac{2}{\varepsilon(1 + (\tau_* E_Z^*)^2)}\right]^2 
\left[\frac{1}{4}(\tau_*\tilde{V})^2 + \frac{\pi^2}{3} (\tau_* T)^2\right].
\ee
Using the results (\ref{j1mt1}) and (\ref{j2mt1}) in \req{ja} and then
\req{dgmtj}, we arrive at \reqs{poo}.

At high voltages $\tilde{V}\gg T$ (but still low temperatures $T\ll 1/\tau_*$),
we can approximate the hyperbolic cotangents in \reqs{Jmt1}-(\ref{Jmt2})
with their zero-temperature limit: $\coth \frac{\omega}{2T} \to 
\mathrm{sgn} \, \omega$. Performing the differentiation with respect to $\tilde{V}$
and a partial integration, and using the approximation in \req{arc},
we obtain
\bes\label{j1int}
J^1_{MT} =
&\frac{1}{2\varepsilon^2} \sum_{n,m=\pm 1}\int_{-\tilde{V}}^{\tilde{V}}\!\!\!\!d\omega \, \frac{\tau^*\arctan{\tau^*(\omega + \tilde{V} - n E^*_Z)}}{1+\tau_*^2(\omega - \tilde{V} - m E^*_Z)^2} .
\end{split}\ee 
The result of the  $\omega$-integral can be expressed in terms of dilogarithms:
\bea
J^1_{MT}&=&\frac{ 1}{2\varepsilon^2}\sum_{m=\pm 1} h(\tau^*\tilde{V},\tau^*\tilde{V}+m \tau^*E_Z^*),\\
J^2_{MT}&= &\frac{1}{2}J_1^{MT}-\frac{ 1}{4\varepsilon^2}\left(\sum_{m=\pm 1} \arctan{\tau^*(\tilde{V}-mE_Z^*)}\right)^2,\nonumber
\eea
The second line is found in the approximation (\ref{arc}) by direct comparison 
of the definitions (\ref{Jmt1})-(\ref{Jmt2}) and the function $h$ is defined 
as:
\be\label{hdef}
h(x,y) \equiv 
\frac{1}{2}\Re\left\{f(x,y) - f(-x,y)+f(y,x)-f(-y,x)\right\}\! ,
\ee 
with
\bes
&f(x,y) \equiv  \Li\left(\frac{x+y+i}{2(y-i)}\right) -\Li\left(\frac{x+y+i}{2y}\right)\\
&+\ln{\left(1+ix+iy\right)}\left[\ln{\frac{y-x-i}{2(y-i)}}-\ln{\frac{y-x+i}{2y}}\right].
\end{split}
\ee 
The approximate expression (\ref{approxh}) for $h$ is found as follows:
we first notice that in \req{j1int} we can restrict the integral to the 
region $(0,\tilde{V})$ (if we multiply by 2); then we shift the integration variable:
$\omega \to \omega+\tilde{V}+mE_Z^*$ and we finally neglect the dependence of the
numerator on $\omega$. The resulting integral is straightforward and gives the 
formula (\ref{approxh}).

We now turn to the high temperature regime $T \gg \frac{1}{\tau_*} $. 
In this case, we can neglect $1/(4\pi\tau_* T)$ in the argument of the digamma 
function in the definition (\ref{khay}); then using the identity
\be\label{impsi}
\frac{2}{\pi}\Im\, \psi^{(0)}\left(\frac{1}{2}+i \frac{\omega}{\pi}\right) = 
\tanh\omega,  \quad\quad (\Im\,\omega = 0),
\ee
we have
\be\label{kisi}
\ipsi{0}{m}{\alpha} \approx \frac{1}{2}\tanh\left(\frac{\omega - \alpha \tilde{V} 
-m E_Z^*}{4T}\right) .
\ee
Substituting the above approximate expression into \reqs{Jmt1}-(\ref{Jmt2}) 
and performing both the differentiation with respect to $\tilde{V}$ and the 
integration over $\omega$ we obtain
\bea
J^1_{MT} &\approx& \frac{1}{2}\frac{\pi^2}{\varepsilon^2}
\Bigg\{ 1 - \frac{1}{2}\coth{\left(\frac{E_Z^*}{2T}\right)} \\
&& \times\left[c_1\left(\frac{E_Z^* - \tilde{V}}{2T}\right) + c_1\left(\frac{E_Z^*+\tilde{V}}{2T}\right) \right]\!\Bigg\},
\nonumber
\\
J^2_{MT} &\approx& \frac{\pi^2}{\varepsilon^2} \sum_{m=\pm 1} \frac{1}{8} c_2\left(\frac{\tilde{V}-mE_Z^*}{2T}\right),
\eea
where
\be\label{cdef}
c_n(x)\equiv \frac{d^n}{dx^n}(x\coth{x}) \, .
\ee
Since $J_{DoS_{1}}$ is smaller than $J_{MT}$ by a factor of $(\tau_* T)^{-1}$ [see \reqs{sonbol}], it can always be neglected in this regime.

This concludes the calculation of the Maki-Thompson and ``$\Re \, \C^2$'' 
contribution and hence the derivation of the results for $\Delta G_{MT}$ 
reported in Sec.~\ref{sec:results}. In the next subsection we present
the calculation of the remaining ``DoS'' correction for 
completeness.

\subsubsection{DoS$_{2}$ part}

In this subsection we evaluate $J_{Dos_{2}}$, \req{dooos}, in the relevant 
limits.
Using \reqs{delf}, (\ref{gg2}) and (\ref{cooperon}),
integration over $\epsilon$ can be done similarly to \req{kallepache} by closing the integration contour in the upper half plane. The result is
\bea\label{jdos11}
J_{DoS_{2}} &=& \frac{1}{16\tau_* T}\!\!\sum_{m,\alpha,\beta}\!\!\beta \rmg_\alpha   \frac{\partial}{\partial \tilde{V}} \!\int\!\frac{d\omega}{2\pi}\coh{2\omega \!-\!(\alpha+\beta)\tilde{V}} \nonumber \\ &
\times&\Re\!\left\{\big[\myxic{}{m}{\beta}(-\omega)-\myxi{}{m}{\alpha}(\omega)
\big]\D_\Delta^R(\omega)\right\}\! ,
\eea
where 
\be
\myxi{}{m}{\alpha}(\omega) \equiv \frac{1}{\pi}\psi^{(1)}\left(\frac{1}{2}+\frac{\frac{1}{\tau_*} + i (\omega-m E_Z^* -\alpha \tilde{V})}{4\pi T}\right),
\ee
and the subscripts $m$, $\alpha$ and $\beta$ are summed over $\pm 1$.
In \req{jdos11}, the approximation (\ref{propapprox}) amounts to neglecting
the imaginary part of the propagator in comparison with its real part, so that
\bes\label{DoS}
J_{DoS_{2}} = \frac{1}{16\tau_* T} \sum_{m=\pm 1} \frac{\partial}{\partial \tilde{V}} \int\!\frac{d\omega}{2\pi}\coth\left(\frac{\omega}{2T}\right) \hspace{1.2cm}\\ \times \Re\left[\myxi{}{m}{+}(\omega) -\myxi{}{m}{-} (\omega) \right]\Re\,\D_\Delta^R(0).
\end{split}\ee

In the low temperature regime $T \ll \frac{1}{\tau_*}$, we can use
the following asymptotic expansion: 
\bea
 \label{appr}\psi^{(1)}(z) &\approx& \frac{1}{z} \, , \quad\quad z \gg 1
 \eea
so that
\bes
J_{DoS_{2}} \approx & \frac{1}{4\tau_*} \Re \D_\Delta^R(0)
 \Re\sum_{\alpha,m} \int\!\frac{d\omega}{2\pi}
\coth\left(\frac{\omega}{2T}\right)\\
&\times \frac{\partial}{\partial \tilde{V}} \left[\frac{\alpha}{\frac{1}{\tau_*} +i(\omega -\alpha \tilde{V} - mE_Z^*)}\right].
\end{split}\ee
The result of  the integral can be expressed in terms of $\psi^{(1)}$ and using 
again the expansion (\ref{appr}) we find
 \be
 J_{DoS_{2}} \approx \frac{1}{2\varepsilon}   \sum_{m=\pm 1} \frac{1}{1+\tau_*^2(\tilde{V}-m E_Z^*)^2}.
\ee

In the high temperature limit $T \gg \frac{1}{\tau_*}$, we can use \req{kisi}
to rewrite $ J_{DoS_{2}}$ as
\bea
&& J_{DoS_{2}} \approx  \frac{\Re  \D_\Delta^R(0)}{16\tau_* T} \sum_{m=\pm 1} 
\frac{\partial}{\partial \tilde{V}} \int\!\frac{d\tilde{\omega}}{2}\coth{2\tilde{\omega}} \\ && \times\left[\mathrm{sech}^2(\tilde{\omega}-\tilde{V}-m\tilde{E}_Z)-\mathrm{sech}^2(\tilde{\omega}+\tilde{V}-m\tilde{E}_Z)\right], \nonumber
\eea
and integrating over $\omega$ we obtain
\be
J_{DoS_{2}} \approx \frac{\pi/\varepsilon}{16\tau_* T}  \sum_{m=\pm 1} c_2\left(\frac{\tilde{V}-mE_Z^*}{2T}\right),
\ee
where $c_2$ was defined in \req{cdef}. We see that for $\tau_* T \simeq 1$, we have $J_{DoS_{2}} \approx \varepsilon J_{MT} \gg |J_{MT}|$. However the prefactor $A$ in \req{khaar} vanishes for reflectionless contacts and for this reason we only report the MT contribution in Sec.~\ref{sec:results}.

\section{Conclusions}

In this paper we presented a quantum kinetic equation description of transport
in open quantum dots with the inclusion of all the ``universal'' --
in the Random Matrix Theory sense -- interaction effects [see \req{hint}]. 
While the effect of the charging energy was taken into account 
before,\cite{brouwer:aleiner:99,golubev:zaikin:04,brouwer:lamackraft:05} 
the interaction corrections in the triplet and Cooper channels are 
considered here for the first time.

The main result of the present work is that
the triplet channel interaction can significantly affect the 
differential conductance of the quantum dot, see Fig.~\ref{fig1}, and in 
contrast with the singlet channel contribution, it is sensitive to the 
magnetic field, cf. Eqs.~(\ref{condu}) and (\ref{tripl}) -- see also
Fig.~\ref{fig2}. For transparent contacts
both the singlet and triplet channel corrections to the electrical 
conductance vanish; the non-vanishing Cooper channel contribution
is unfortunately expected to be a negligible one in metallic dots, as 
discussed after \req{redtemp}.

In addition to the electrical conductance, we were able to calculate
the thermal conductance by applying the {\it local} kinetic equation
approach developed in Ref.~\onlinecite{catelani:aleiner:05}. For the thermal conductance we find 
that the Wiedemann-Franz law is violated by the interaction corrections [see \req{thcores}], and we investigated the effect of magnetic field on the Lorentz ratio for contacts of finite reflection. The charge and triplet channel corrections to the thermal conductance also vanish for reflectionless contacts, and the Wiedemann-Franz law  is not violated by the electron-hole channels in this case. 

\acknowledgments

We would like to thank P.W. Brouwer, L.I. Glazman and A.I. Larkin for 
interesting discussions. 

\appendix

\section{Vanishing of $\St_-$ in the steady state}\label{app:stm}

We want to show that the properties (\ref{stmpr}) and (\ref{stmpr2}) 
hold in the steady state. To this end, we notice that the second term on the 
right hand side of the definition (\ref{stmdef}) always satisfies 
these properties, as one can verify using the same arguments that prove the 
corresponding properties (\ref{genprop}) and (\ref{stppr2}) of the collision 
integral $\St_+$. Next we want to show that in the steady state
\be\label{toprove}
\left[ \delta Q \right] K_- = 0,
\ee
after the average over the fluctuating field but before any limit and/or time 
derivative is taken. Indeed in $\delta Q$ only the contributions proportional 
to $\delta g^K_+$ should be kept, since the ones proportional to $\delta g^K_-$
vanish after the average. 
We note that the right hand side of \req{dgkpeq} can be rewritten as
\be
i\left[ K_+ ; \partial_t g^K \right],
\ee
by using the zeroth order (in the interaction) part of \req{dkineq} for $g^K$,
i.e. neglecting the collision integrals $\St_1$ and $\St_2$
-- this is sufficient in the one-loop approximation.
In the steady state, $\partial_t g=0$ and 
hence $\delta g_+=0$. This proves \req{toprove} and therefore the 
properties (\ref{stmpr}) and (\ref{stmpr2}).

\section{Thermodynamics}\label{app:thermodyn}

\begin{figure}[!th]
\begin{center}\includegraphics[width=0.46\textwidth]{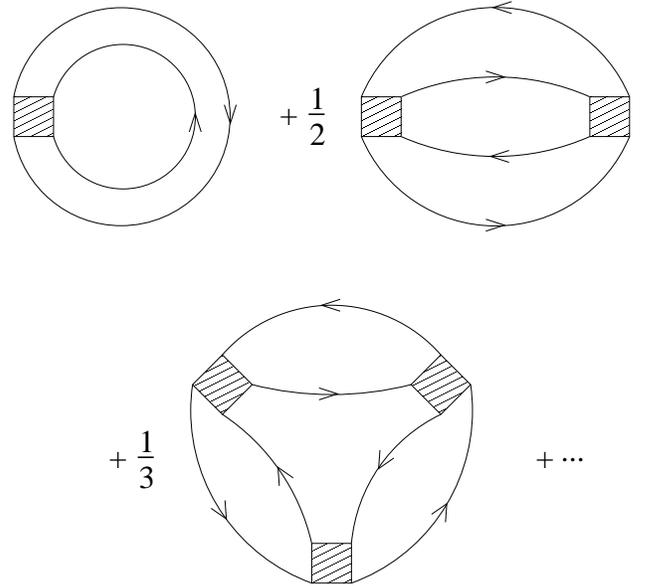}\end{center}
\caption{Leading singular contribution to the thermodynamic
potential. The shaded box corresponds to $2 E_c$, defined through the 
two particle vertex [see e.g. Ref.~\onlinecite{abrikosov}]; 
the solid lines are coherent parts of the 
electron Green functions.}
\label{ring}
\end{figure}

The effects of interaction on the thermodynamics can be obtained by calculating
the leading contribution to the themodynamic potential $\Omega$; this is 
given by the sum of the so-called ring diagrams of Fig.~\ref{ring} [see e.g. 
Ref.~\onlinecite{altshuler:aronov:book}]. This sum can be calculated in the 
Matsubara representation and after 
standard analytic contnuation the correction $\delta\Omega$ is found to be:
\be
\delta\Omega = \int\!\frac{d\omega}{2\pi} \, 
\frac{1}{2}\coth\left(\frac{\omega}{2T}\right) \Im \, 
\ln \Big( 1 - 2 E_c \Pi^R_\phi (\omega) \Big).
\ee
Using expression (\ref{polft}) for the polarization operator and the 
definitions (\ref{diffdef}) and (\ref{Lc}), we can rewrite this as:
\be
\delta\Omega = \int\!\frac{d\omega}{2\pi} \,
\frac{1}{2}\coth\left(\frac{\omega}{2T}\right) \Im \, 
\Big[ \ln {\cal L}^g - \ln {\cal L}^\rho \Big].
\ee

We can now calculate the correction $\delta c_V$ to the specific heat and 
hence arrive at the expression for the energy density $u_b$; indeed
\be\label{cvrel}
\delta c_V = - T \frac{\partial^2 \delta\Omega}{\partial T^2} = 
\frac{\partial u_b}{\partial T}.
\ee
Using the first of the above relations and after an integration by parts
we find
\be\label{cv}
\delta c_V = \frac{\partial}{\partial T} \int_0^\infty\!\!\!d\omega \, \omega
N_P(\omega) \Big[ b^\rho (\omega) - b^g(\omega) \Big],
\ee
where we introduced the bosonic density of states:
\be
b^\alpha(\omega) = \frac{1}{\pi} \Im \, \partial_\omega \ln {\cal L}^\alpha.
\ee
Comparing \req{cv} to the second relation in \req{cvrel} and generalizing 
this result to an arbitrary distribution function we have
\be
u_b=\int_0^\infty\!\!\!d\omega \, \omega
\Big[ N^\rho(\omega) b^\rho (\omega) - N^g(\omega) b^g(\omega) \Big].
\ee
It is straightforward to prove that this coincides with 
\req{ubdef} [by substituing in the latter \req{bosdistfunc} and noticing the
different limits of integration].

\end{document}